\documentclass[12pt]{article}
\usepackage{mathrsfs}
\usepackage{graphicx}
\usepackage{amssymb,amsmath,amsthm}
\usepackage{epstopdf}
\usepackage{bm}
\usepackage{feynmf}
 \textwidth = 6.5 in \textheight = 9 in \oddsidemargin =
0.0 in \evensidemargin = 0.0 in \topmargin = 0.0 in \headheight =
0.0 in \headsep = 0.0 in \voffset=0.0 in
\def\ba{\begin{eqnarray}}
\def\ea{\end{eqnarray}}
\def\nn{\nonumber}

\def\D{\mathcal{D}}

\def\:{\boldsymbol{:}}

\def\Tr{\textrm{Tr}}

\long\def\symbolfootnote[#1]#2{\begingroup%
\def\thefootnote{\fnsymbol{footnote}}\footnote[#1]{#2}\endgroup}

\begin{document}

\begin{flushright}
\end{flushright}
\begin{center} 
\vglue .06in
{\Large \bf {A Kaluza-Klein Model with Spontaneous Symmetry Breaking: Light-Particle Effective Action and Its Compactification Scale Dependence\symbolfootnote[1]{This work was supported by the US department of energy.}}}\\[.5in]

{\bf Ratindranath Akhoury\symbolfootnote[2]{Electronic address: akhoury@umich.edu} and Christopher S. Gauthier\symbolfootnote[3]{Electronic address: csg@umich.edu}}\\
[.1in]
{\it{Michigan Center for Theoretical Physics\\
Randall Laboratory of Physics\\
University of Michigan\\
Ann Arbor, Michigan 48109-1120, USA}}\\[.2in]
\end{center}

\begin{abstract}
We investigate decoupling of heavy Kaluza-Klein modes in an Abelian Higgs model with space-time topologies $\mathbb{R}^{3,1} \times S^{1}$ and $\mathbb{R}^{3,1} \times S^{1}/\mathbb{Z}_{2}$. After integrating out heavy KK modes we find the effective action for the zero mode fields. We find that in the $\mathbb{R}^{3,1} \times S^{1}$ topology the heavy modes do not decouple in the effective action, due to the zero mode of the 5-th component of the 5-d gauge field $A_{5}$. Because $A_{5}$ is a scalar under 4-d Lorentz transformations, there is no gauge symmetry protecting it from getting mass and $A_{5}^{4}$ interaction terms after loop corrections. In addition, after symmetry breaking, we find new divergences in the $A_{5}$ mass that did not appear in the symmetric phase. The new divergences are traced back to the gauge-goldstone mixing that occurs after symmetry breaking. The relevance of these new divergences to Symanzik's theorem is discussed. In order to get a more sensible theory we investigate the $S^{1}/\mathbb{Z}_{2}$ compactification. With this kind of compact topology, the $A_{5}$ zero mode disappears. With no $A_{5}$, there are no new divergences and the heavy modes decouple. We also discuss the dependence of the couplings and masses on the compactification scale. We derive a set of RG-like equations for the running of the effective couplings with respect to the compactification scale. It is found that magnitudes of both couplings decrease as the scale $M$ increases. The effective masses are also shown to decrease with increasing compactification scale. All of this opens up the possibility of placing constraints on the size of extra dimensions.
\end{abstract}

\section{Introduction}
The possibility that our world may contain more then the usual four space-time dimensions is an idea that has captivated physicists for almost a century now. Extra-dimensional theories have arisen in a variety of areas to explain a wide range of phenomena. Today they are most notably used to eliminate the conformal anomaly in string theory. Although there have been several models in particle physics and cosmology that have incorporated extra dimensions, there is still no experimental evidence to suggest their existence. The traditional explanation for why these extra dimensions have so far escaped detection was put forth by Oskar Klein who surmised that these extra dimensions are compactified to such a small scale that they can not be detected by present day accelerators.

A common feature of all Kaluza-Klein models is the existence of an infinite ``tower'' of progressively massive particles. In general Kaluza-Klein models the mass of the lightest heavy KK mode is $M \sim (\textrm{Size of compact dimenions})^{-1}$. If $M$ is higher then the energy scale under consideration, the effective field theory will be a theory of the zero KK mode fields in four dimensions. If the theory is to be consistent with observation, the effects of the heavy KK modes must disappear in the low-energy limit of the theory. Their only effect on the low-energy dynamics are in the form of loop corrections to the local operators of the zero KK mode fields, which are then absorbed into the existing couplings, masses, and field redefinitions of the tree-level theory \cite{Appelquist:1974tg}. When this happens the heavy modes are said to have decoupled from the low-energy physics.

There have been a number of works that study certain aspects of the quantum effective action of a theory with small extra dimensions \cite{Appelquist:1983vs,RandjbarDaemi:2006gf,DiClemente:2001ge} and their phenomenological implications \cite{Antoniadis:1990ew,Appelquist:2000nn}. The compactification scale dependence of the low-energy effective action has been studied previously in different models. For example, in a $\phi^{4}$ model in arbitrary spacetime dimensions \cite{Sochichiu:1999ih,Sochichiu:1999mk}, and in scalar QED in 5-D \cite{Akhoury:2007dz}. In distinction to earlier works, this paper will address the problem of heavy KK mode decoupling at the level of the effective action in a theory with spontaneous symmetry breaking. In particular we will investigate heavy mode decoupling in the Kaluza-Klein extension of the Abelian Higgs model. We focus our attention on the {\it light particle effective action} (LPEA). This is the quantum effective action obtained after integrating out only those fields with masses greater than some predetermined scale. The result is a low-energy effective theory of light fields with masses and couplings modified by the heavy loop corrections. In our analysis, the energy scale under consideration is assumed to be around the electroweak scale ($\sim$ TeV), rendering quantum gravitational effects unimportant. The scale is lower then the lightest heavy mode, therefore the LPEA will consist of only the zero KK mode fields. We will assume that space-time is 4+1 dimensional, with an extra spatial dimension that has been periodically identified: $x^{5} \sim x^{5} + 2 \pi R$, where $R$ is the radius of the $5$-th dimension. The resulting space-time manifold has a $\mathbb{R}^{3,1} \times S^{1}$ topology. Later in the paper we will also consider the case of $\mathbb{R}^{3,1} \times S^{1}/\mathbb{Z}_{2}$ orbifold topology.

In the Kaluza-Klein theory with scalars, the structure of the compact dimensions, requires that the fields have periodic boundary conditions along the compact directions. This allows us to represent the 5-d scalar field as a Fourier series of 4-d field modes, whose fundamental frequency is proportional to $M=R^{-1}$. After integrating over the fifth dimension, the 5-d action becomes the 4-d action of a light field sector containing the zero KK modes, a heavy sector containing the nonzero KK modes and an interaction term that connects the two:
\begin{gather}
S^{(\textrm{5d})}
\rightarrow
S_{0}^{(\textrm{4d})}
+
\sum_{n=1}^{\infty}
S_{n}^{(\textrm{4d})}
+
S_{\textrm{int}}^{(\textrm{4d})}.
\end{gather}
At energies below $M$ the heavy modes decouple. The interaction term creates Feynman diagrams with heavy mode loops that make corrections to the masses and couplings of the light sector. These corrections will come in the form of an integral over the  4-d dimensional momentum and a sum over the heavy KK modes:
\begin{gather}
\sum_{n=1}^{\infty}
\int \frac{d^{4} p}{(2 \pi)^{4}} f_{n}(p).
\label{GeneralExpression}
\end{gather}
In general these quantities are divergent, and a scheme for regularizing and subtracting the divergences must be defined. Many procedures have been suggested for regularizing expressions similar to (\ref{GeneralExpression}) \cite{GrootNibbelink:2001bx}, \cite{DiClemente:2001ge,Contino:2001uf}, \cite{Delgado:2001ex}, \cite{Contino2:2001gz}, \cite{Ghilencea:2001ug,Ghilencea:2001bv,Ghilencea:2001bw}. There is a ongoing debate as to which procedure makes the most sense physically \cite{Alvarez:2006sf}. We do not attempt to answer this question, and we will simply choose the procedure that makes the most physical sense to us. In the context of our discussion, the corrections are treated as a sum of 4-d loop corrections originating from an infinite tower of massive particles. Therefore, we choose to evaluate the 4-d integral first using dimensional regularization, and then performing the sum over KK modes using zeta function regularization \cite{DiClemente:2001ge}. 

The extension of Kaluza-Klein theories to gauge fields is similar to scalar fields, but with some additional subtleties. The first of these is gauge fixing in 5-d Kaluza-Klein models. It is a straightforward task to generalize the gauge fixing done in four dimensions to five dimensions \cite{Dienes:1998vg,Muck:2001yv,Papavassiliou:2001be}. The biggest difference between scalar and gauge fields in Kaluza-Klein models is the additional scalar one obtains after dimensional reduction. The extra component $A_{5}$ of the 5-d gauge field becomes a KK tower of 4-d Lorentz scalars after dimensional reduction. Being unprotected by any gauge symmetry in 4-d, corrections by heavy modes can create new local operators involving the $A_{5}$ zero mode that do not originate from any local operator in the 5-d theory. 

The primary focus of this paper will be on the effects spontaneous symmetry breaking has on decoupling in Kaluza-Klein models. It is well known that in 3+1 space-time dimensions, the gauge field in an Abelian Higgs gains a mass after symmetry breaking by ``eating'' the would-be (or pseudo) Goldstone boson \cite{Higgs:1966ev}. In a sense the gauge and Goldstone bosons are mixed together. The renormalizability of this theory can be shown by appealing to a result from Symanzik \cite{Symanzik:1969ek,Symanzik:1970}, which tells us that the divergences in the unbroken phase where the gauge field is massless are the same as those in the broken symmetry phase. Since the theory of a massless gauge boson is known to be renormalizable then it follows from Symanzik that the theory is also renormalizable in the broken phase. This is a result that we wish to have carry over to the Kaluza-Klein version of the Higgs mechanism.

This picture remains largely intact in the 5-d Kaluza-Klein extension, only now we have the extra 4-d scalar $A_{5}$ to deal with. To help us understand how this will effect the low-energy dynamics we note that the dimensionally reduced action can be divided into two sectors. The first sector contains the 4-d gauge field modes and their interactions with themselves and the Higgs and Goldstone fields. The second sector contains the $A_{5}$ modes and their interactions. These two sectors only communicate through their shared interaction with the Higgs and Goldstone fields.

Problems arise when one tries to calculate loop corrections in the $A_{5}$ sector after symmetry breaking. In the gauge sector the Goldstone has mixed with the gauge field to become its longitudinal component, and no longer exists as a physical degree of freedom. The Goldstone's effects through loop corrections to the gauge and Higgs fields are canceled due to local $U(1)$ gauge symmetry, and as a result, no new divergences are introduced. However, in the $A_{5}$ sector, no such symmetry protects the field from getting nonzero corrections from the would-be Goldstone. This presents a potential problem, since the would-be Goldstone can now introduce new divergences that require we use different counter terms than in the symmetric phase.

Assuming that any problems with the $A_{5}$ mode can be dealt with, there is the additional question of the compactification scale's effect on the LPEA. Although the heavy modes have been integrated out at low energies, their presence is still felt by the light fields, whose couplings have been modified by an additive factor proportional to $\log M$. In the conventional MS scheme the $\log \mu$ piece can be used to derive the renormalization group equation for a coupling. This same procedure can be used to construct a set of RG-like equations for the running of the effective couplings with respect to the scale $M$. The resulting solutions are the renormalization group improvement of the one-loop corrected couplings. With such solutions we can better hypothesize on the effective action's dependence on the size of hidden dimensions.

For a review of the LPEA see refs \cite{Weisberger:1981xe}. In section \ref{5dAbelianHiggsTitle} we derive the LPEA of the 5-d Abelian Higgs model. Here the issues concerning gauge fixing and Ward identities are discussed in detail. In section \ref{MassandCouplingRenormalization} we talk about the subtraction of divergences from the LPEA. It is found that the LPEA of this model without the $A_{5}$ zero mode can be made finite, and a set of equations for running of the scalar and gauge couplings with respect to the compactification scale are derived. We also discuss the failure of the symmetric phase counter terms to absorb the infinite corrections of the theory in the broken phase. Furthermore, it is pointed out that a local $A_{5}^{4}$ operator is introduced through one-loop corrections, explicitly violating decoupling. A resolution to the problem of new divergences is discussed in section \ref{CircleOrbifoldCpct}. It is shown that no additional divergences are created in the broken phase if an orbifold compactification is used. In addition, the heavy modes, which did not decouple in the circle compactification, do decouple in the orbifold case. Finally, in section \ref{Conclusions} we conclude our paper with a summary of our main results. In the appendices we outline the procedure used to derive the LPEA, and use it to find the low-energy effective action for massive $\phi^{4}$ theory with and without SSB. Here we also show explicitly the results from the zeta function regularization of common divergent sums used in the course of this study. 

\section{5-d Abelian Higgs Model}
\label{5dAbelianHiggsTitle}
Consider the five dimensional action of an Abelian gauge field $\bar{A}_{M}$ minimally coupled to a scalar field $\bar{\phi}$\footnote{In this paper a bar over fields and coordinates denotes the $4+1$ dimension fields and coordinates, while those without the bar denotes their dimensionally reduced counterparts. Indices $M,N,...,$ etc. take values over the total number of compact and non-compact dimensions while $\mu,\nu,..,$ etc. will denote indices over the non-compact dimensions.}:
\begin{gather}
S
=
\int d^{5}\bar{x}
\left[
-\frac{\bar{Z_{A}}}{4}
\bar{F}_{MN}
\bar{F}^{MN}
+ 
\bar{Z}_{\phi}
|\D_{M} \bar{\phi}|^{2}
-
\bar{Z}_{\phi}
\bar{m}_{b}^{2} |\bar{\phi}|^{2}
-
\frac{\bar{Z}_{\phi}^{2} \bar{\lambda}_{b}}{3!}
|\bar{\phi}|^{4}
\right].
\label{AbelianHIggsAction}
\end{gather}
The covariant derivative is given by $\D_{M} = \partial_{M} + i \bar{e}_{b} \bar{Z}_{A}^{1/2} \bar{A}_{M}$. In this model $\bar{\phi}$ is complex and has a charge $\bar{e}$ that couples to a $U(1)$ gauge field $\bar{A}_{M}$. The couplings $\bar{\lambda}_{b}$, $\bar{m}_{b}^{2}$ and $\bar{e}_{b}$ are the bare couplings and we express them in terms of their physical couplings and counter terms like so: 
\begin{gather}
\bar{\lambda}_{b} = \bar{\lambda} ( 1 + \bar{ \delta}_{\lambda}),
\qquad
\bar{e}_{b} = \bar{e} ( 1 +  \bar{\delta}_{e}),
\qquad
\bar{m}_{b}^{2} = \bar{m}^{2} ( 1 + \bar{\delta}_{m^{2}}).
\end{gather}
Here, $\bar{\lambda}$, $\bar{m}^{2}$, and $\bar{e}$ are the physical couplings and $\bar{ \delta}_{\lambda}$, $\bar{ \delta}_{m^{2}}$, and $\bar{ \delta}_{e}$ are their counter terms. We assume that $\bar{m}^{2} <0$, and therefore the classical vacuum breaks the local $U(1)$ gauge  symmetry. Expanding the scalar $\bar{\phi}$ around the VEV at $|\bar{\phi}| = \frac{\bar{v}}{\sqrt{2}}$ breaks the $U(1)$ symmetry explicitly and generates masses for the gauge and Higgs fields. Expanding the Higgs field around its VEV $\bar{v} = \sqrt{- \frac{6 \bar{m}^{2}}{\bar{\lambda}}}$, the action (\ref{AbelianHIggsAction}) becomes
\begin{gather}
S
=
\int d^{5}\bar{x}
\left[
-
\frac{\bar{Z}_{A}}{4}
\bar{F}_{MN}
\bar{F}^{MN}
- 
\bar{Z}_{v}^{1/2} \bar{Z}_{\phi} \bar{Z}_{A}^{1/2} \bar{e}_{b}\bar{v}
\bar{A}_{M} \partial^{M} \bar{\varphi}
+
\frac{\bar{Z}_{\phi}}{2}
\partial_{M} \bar{\varphi}
\partial^{M} \bar{\varphi}
+
\frac{\bar{Z}_{\phi}}{2}
\partial_{M} \bar{\chi}
\partial^{M} \bar{\chi}
\right.
\nn \\
+ 
\bar{Z}_{\phi} \bar{Z}_{A}^{1/2} 
\bar{e}_{b} \bar{A}_{M}
(\bar{\varphi} \partial^{M} \bar{\chi} 
-
\bar{\chi} \partial^{M} \bar{\varphi} )
+
\frac{\bar{Z}_{\phi} \bar{Z}_{A} \bar{e}_{b}^{2}}{2} 
(\bar{\varphi}^{2} + \bar{\chi}^{2}) \bar{A}_{M} \bar{A}^{M}
+
 \bar{Z}_{v}^{1/2} \bar{Z}_{\phi} \bar{Z}_{A}
\bar{e}_{b}^{2} \bar{v}
\bar{\chi} \bar{A}_{M} \bar{A}^{M}
\nn \\
+
\frac{\bar{Z}_{v} \bar{Z}_{\phi} \bar{Z}_{A} \bar{e}_{b}^{2} \bar{v}^{2}}{2}
\bar{A}_{M} \bar{A}^{M}
-
\frac{\bar{Z}_{v}^{1/2}  \bar{Z}_{\phi}}{6}
(\bar{Z}_{v} \bar{Z}_{\phi} \bar{\lambda}_{b} \bar{v}^{2} + 6 \bar{m}_{b}^{2})
\bar{v}
\bar{\chi}
-
\frac{\bar{Z}_{\phi}}{12} (\bar{Z}_{v} \bar{Z}_{\phi} \bar{\lambda}_{b} \bar{v}^{2} + 6 \bar{m}_{b}^{2}) 
\bar{\varphi}^{2}
\nn \\
\left.
-
\frac{\bar{Z}_{\phi}}{4}
(\bar{Z}_{v} \bar{Z}_{\phi} \bar{\lambda}_{b} \bar{v}^{2} + 2 \bar{m}_{b}^{2}) \bar{\chi}^{2}
-
\frac{\bar{Z}_{v}^{1/2} \bar{Z}_{\phi}^{2} \bar{\lambda}_{b} \bar{v}}{3!}
\bar{\chi}
(\bar{\varphi}^{2} + \bar{\chi}^{2})
-
\frac{ \bar{Z}_{\phi}^{2} \bar{\lambda}_{b}}{4!}
(\bar{\varphi}^{2} + \bar{\chi}^{2})^{2}
\right].
\label{5DAction}
\end{gather}
Be aware that we have included an additional counter term $\bar{Z}_{v}$. This counter term is introduced as a reshifting of the VEV; $\bar{v} \rightarrow \bar{Z}_{v}^{1/2} \bar{v}$. It is needed to absorb an additional divergence that may be generated by gauge fixing \cite{Collins:1984xc}. Although not always needed, in the gauge we will be working in, $\bar{Z}_{v}$ will differ from unity.

For any bosonic field $\bar{\Phi}$ to be well defined on the space-time manifold $\mathbb{R}^{3,1} \times S^{1}$, the field must obey periodic boundary conditions in the compact direction: $\bar{\Phi}(x^{\mu} , x^{5}) = \bar{\Phi}(x^{\mu},x^{5} + 2 \pi R)$. Since $\bar{\Phi}$ satisfies periodic boundary conditions, it can be expressed as a Fourier series in the compact direction:
\begin{gather}
\bar{\Phi}(\bar{x})
=
\sum_{n=-\infty}^{\infty} \Phi^{(n)}(x) e^{i n M \theta}
\end{gather}
where $\theta$ is an angular coordinate patch on the compact dimension and $M = R^{-1}$ is the compactification scale. The action of this field can now be reduced to a 4-d action by integrating over the compact dimension. The result is an action for an infinite number of coupled fields $\Phi^{(n)}$, indexed by the magnitude of their compact momenta. After integrating out the 5-th direction, the 5-d tree-level action (\ref{5DAction}) becomes:
\begin{gather}
S
=
\int d^{4}x
\Bigg[
\sum_{n = -\infty}^{\infty}
\left(
\frac{1}{2}
A_{\mu}^{(n)}
[
g^{\mu\nu}
[
\square_{4}
+
m_{A}^{2}
+
n^{2} M^{2}
]
-
\partial^{\mu} \partial^{\nu}
]A_{\nu}^{(-n)}
-
\frac{1}{2} A_{5}^{(n)}
\left[
\square_{4}
+
m_{A}^{2} 
\right]
A_{5}^{(-n)}
\right.
\nn \\
-
i n M
A_{\mu}^{(n)}
\partial^{\mu}
A_{5}^{(-n)}
-
m_{A} 
A_{\mu}^{(n)} \partial^{\mu} \varphi^{(-n)}
-
i n M m_{A}
A_{5}^{(n)} \varphi^{(-n)}
+
\frac{1}{2} \partial_{\mu} \varphi^{(n)} \partial^{\mu}\varphi^{(-n)}
\nn \\
\left.
+
\frac{1}{2} \partial_{\mu} \chi^{(n)} \partial^{\mu}\chi^{(-n)}
-
\frac{1}{2}  n^{2} M^{2} \varphi^{(n)} \varphi^{(-n)} 
-
\frac{1}{2} \left[m_{\chi}^{2} + n^{2} M^{2}\right] \chi^{(n)} \chi^{(-n)} 
\right)
\nn \\
+
\sum_{k,l=-\infty}^{\infty}
\left[
e
A_{\mu}^{(k)}
\left(
 \varphi^{(l)} \partial^{\mu}\chi^{(-k-l)}
-
 \chi^{(l)} \partial^{\mu}\varphi^{(-k-l)}
 \right)
+
i e (k+ l) M
 A_{5}^{(k)} 
\left( 
\varphi^{(l)} \chi^{(-k-l)}
-
\chi^{(l)} \varphi^{(-k-l)}
\right)\right.
\nn \\
\left.
+
e m_{A} 
\chi^{(k)} 
\left(
A_{\mu}^{(l)} A^{\mu (-k-l)}
-
A_{5}^{(l)} A_{5}^{(-k-l)}
\right)
-
\frac{\eta}{3!}
(\varphi^{(k)} \varphi^{(l)} + \chi^{(k)} \chi^{(l)})\chi^{(-k-l)}
\right]
\nn \\
+
\sum_{k,l,n=-\infty}^{\infty}
\left(
\frac{e^{2}}{2}
\left(
\varphi^{(k)} \varphi^{(l)} 
+
\chi^{(k)} \chi^{(l)} 
\right)
\left(
A_{\mu}^{(n)} A^{\mu (-k-l-n)}
-
A_{5}^{(n)} A_{5}^{(-k-l-n)}
\right)
\right.
\nn \\
\left.
\left.
-
\frac{\lambda}{4!}
(\varphi^{(k)} \varphi^{(l)} \varphi^{(n)} \varphi^{(-k-l-n)} + 2 \chi^{(k)} \chi^{(l)} \varphi^{(n)} \varphi^{(-k-l-n)} + \chi^{(k)} \chi^{(l)} \chi^{(n)} \chi^{(-k-l-n)} )
\right)
\right],
\label{OriginalDRedAction}
\end{gather}
where
\begin{gather}
m_{\chi}^{2} = \frac{\bar{\lambda} \bar{v}^{2}}{3}
,
\qquad
m_{A}^{2} = \bar{e}^{2} \bar{v}^{2} 
,
\qquad
v = \sqrt{2 \pi R} \, \bar{v},
\nn \\
e = \frac{\bar{e}}{\sqrt{2 \pi R}}
,
\qquad
\eta = \frac{\bar{\lambda} \bar{v}}{\sqrt{2 \pi R}}
,
\qquad
\lambda = \frac{\bar{\lambda}}{2 \pi R}.
\end{gather}
The relations between the masses and the couplings $\bar{\lambda}$ and $\bar{e}$ will hold to all orders of perturbation theory. Since the heavy mode sector will be integrated out in the LPEA then the only relevant counter terms will be those involving the light sector fields only. The counter term action is 
\begin{gather}
\delta S
=
\int d^{5}\bar{x}
\left[
-
\frac{\delta_{A}}{4}
F_{\mu\nu}
F^{\mu\nu}
+
\frac{m_{A}^{2}}{2}
(2 \delta_{e} + \delta_{v} +  \delta_{\phi}  + \delta_{A})
A_{\mu} A^{\mu}
+
\frac{\delta_{5}}{2}
\partial_{\mu} A_{5}
\partial^{\mu} A_{5}
\right.
\nn \\
-
\frac{m_{A}^{2}}{2}
(2 \delta_{e} + \delta_{v} +  \delta_{\phi}  + \delta_{A})
A_{5}^{2}
-
m_{A}
(\delta_{e} + \frac{1}{2} \delta_{v} + \delta_{\phi} + \frac{1}{2} \delta_{A}) 
A_{\mu} \partial^{\mu} \varphi
+
\frac{\delta_{\phi}}{2}
\partial_{\mu} \varphi
\partial^{\mu} \varphi
+
\frac{\delta_{\phi}}{2}
\partial_{\mu} \chi
\partial^{\mu} \chi
\nn \\
+
e
(\delta_{e} + \delta_{\phi} + \frac{1}{2}\delta_{A}) 
A_{\mu}
(
\varphi \partial^{\mu} \chi
-
\chi \partial^{\mu} \varphi)
+
\frac{e^{2}}{2}
(2 \delta_{e} +  \delta_{\phi} + \delta_{A})
(\varphi^{2} + \chi^{2}) 
A_{\mu} A^{\mu}
\nn \\
-
\frac{e^{2}}{2}
(2 \delta_{e} +  \delta_{\phi} + \delta_{A})
(\varphi^{2} + \chi^{2}) 
A_{5}^{2}
+
e
m_{A}
( 2 \delta_{e} + \frac{1}{2} \delta_{v} + \delta_{\phi} + \delta_{A})
\chi A_{\mu} A^{\mu}
\nn \\
-
e
m_{A}
( 2 \delta_{e} + \frac{1}{2} \delta_{v} +  \delta_{\phi} + \delta_{A})
\chi A_{5}^{2}
-
\delta \sigma \chi
-
\frac{\delta \sigma}{2 v}
\varphi^{2}
-
\frac{1}{2}
[
m_{\chi}^{2}
(\delta_{\lambda} + \delta_{v} + 2 \delta_{\phi})
+
\frac{\delta \sigma}{v}
]
\chi^{2}
\nn \\
\left.
-
\frac{\eta}{3!}
(\delta_{\lambda} + \frac{1}{2}\delta_{v} + 2 \delta_{\phi})
\chi
(\varphi^{2} + \chi^{2})
-
\frac{\lambda}{4!}
(\delta_{\lambda} + 2 \delta_{\phi})
(\varphi^{2} + \chi^{2})^{2}
\right],
\end{gather}
where 
\begin{gather}
\delta \sigma = \frac{\lambda v^{3}}{6} (\delta_{v} + \delta_{\phi} + \delta_{\lambda}  - \delta_{m^{2}})
,
\qquad
\delta_{\lambda} = \bar{\delta}_{\lambda}
,
\qquad
\delta_{m^{2}}
=
\bar{\delta}_{m^{2}}
,
\qquad
\delta_{e}
=
\bar{\delta}_{e}
.
\end{gather}
In this paper, our major area of concern will focus on the properties of the LPEA and the effect that spontaneous symmetry breaking has on it. Before we can calculate this, we must first deal with the subtle issues of gauge invariance and unitarity in a model with spontaneous symmetry breaking.

\subsection{Gauge Fixing the Action}
The addition of a gauge field and spontaneous symmetry breaking presents new problems to be dealt with, in particular the issue of gauge fixing. The simplest choice of gauge to use is a 5-d generalization of the 't Hooft-Feynman gauge \cite{Muck:2001yv}:
\begin{gather}
\mathscr{L}_{GF}
=
-\frac{1}{2 \xi} (\partial^{\mu} \bar{A}_{\mu} + \kappa_{1} \xi \partial^{5} \bar{A}_{5} + \kappa_{2} \xi m_{A} \bar{\varphi})^{2}.
\end{gather}
If we set $\xi = \kappa_{1} = \kappa_{2} =1$, the job of finding the LPEA is greatly simplified, and hence this choice will be used in this paper. This gauge choice is appealing because it simplifies the $\bar{\varphi}$, $\bar{A}_{\mu}$ and $\bar{A}_{5}$ propagators and eliminates quadratic cross-terms. Additionally, it also gives each KK mode of $\bar{A}_{5}$ a KK mass $n M$ in the dimensionally reduced action. This is a great simplification since it eliminate the nonzero $A_{5}$ KK modes from the low-energy theory. The price paid for this simplification is that the ghosts do not decouple from the rest of the action.

Since the KK modes of $\bar{A}_{5}$ only get a KK mass in a specific gauge, then in a more general gauge the nonzero KK modes will not have a mass, and therefore won't decouple in the low-energy theory. However, in 4-d the gauge transformation acts on the KK modes of $\bar{A}_{5}$ like
\begin{gather}
\delta A_{5}^{(n)}
=
-\frac{ i n}{e} \Lambda^{(n)},
\label{A5Transform}
\end{gather}
where $\Lambda^{(n)}$ is the $n$-th term in the Fourier expansion of the 5-d gauge parameter $\bar{\Lambda}$. Therefore, with the exception of the zero mode (which transforms like $\delta A_{5}^{(0)} = 0$), the $A_{5}$ KK modes are unphysical gauge degrees of freedom. So not only does the Lorentz gauge simplify the 4-d gauge field propagator, it also eliminates the unphysical $A_{5}$ KK modes from the low-energy theory. While the $n\neq 0$ $A_{5}$ modes are of no consequence, $A_{5}^{(0)}$ is a physical degree of freedom and does not decouple in the low-energy theory. We have to contend with corrections to the mass and couplings of $A_{5}^{(0)}$, which in general are not the same as those of the $A_{\mu}$ zero mode. This is not surprising, however, since the former are not protected by the gauge symmetries of the 4-d theory.

If we combine the new gauge fixing term with the terms that are quadratic in the fields $\bar{\varphi}$, $\bar{A}_{\mu}$, and $\bar{A}_{5}$ we find 
\begin{gather}
\frac{1}{2}
\bar{A}_{\mu}
\left[
\bar{g}^{\mu\nu}( (1 + \bar{\delta}_{A})  [\square_{4} - \partial_{5}^{2}] + m_{A}^{2} [1 + 2 \bar{\delta}_{e} + \bar{\delta}_{\phi} +  \bar{\delta}_{A}] )
-
\bar{\delta}_{A}
\partial^{\mu}
\partial^{\nu}
\right]
\bar{A}_{\nu}
\nn \\
-
\frac{1}{2}
\bar{A}_{5}
\left[
\bar{Z}_{A}\square_{4}
-
\partial_{5}^{2}
+
m_{A}^{2} [1 + 2 \bar{\delta}_{e} + \bar{\delta}_{\phi} +  \bar{\delta}_{A}]
\right]
\bar{A}_{5}
-
\frac{\bar{\delta}_{A}}{2}
 \partial_{5} \bar{A}_{5}\partial^{\nu}\bar{A}_{\nu}
-
\frac{\bar{\delta}_{A}}{2}
 \partial_{5}\bar{A}_{\mu} \partial^{\mu} \bar{A}_{5}
\nn \\
-
\frac{1}{2}
\bar{\varphi}
[
(1 + \bar{\delta}_{\phi} )
(\square_{4} 
-
\partial_{5}^{2})
+
m_{A}^{2} + \frac{\bar{\lambda} \bar{v}^{2}}{6} ( \bar{\delta}_{\lambda} + \bar{\delta}_{\phi} - \bar{\delta}_{m^{2}})]
\bar{\varphi}
\nn \\
+
m_{A}(\bar{\delta}_{e} + \bar{\delta}_{\phi} + \frac{1}{2} \bar{\delta}_{A})
\bar{\varphi}
\partial^{\mu}\bar{A}_{\mu}
+
m_{A}(\bar{\delta}_{e} + \bar{\delta}_{\phi} + \frac{1}{2} \bar{\delta}_{A})
\bar{\varphi}
\partial^{5}\bar{A}_{5}.
\end{gather}
Notice the gauge fixing terms have no counter terms themselves. This is a consequence of the Slavnov-Taylor identities, which can be used to show that the gauge fixing parameters need no further subtractions \cite{Collins:1984xc}. As a result, the tree-level cross-terms between the $\bar{A}_{\mu}$, $\bar{A}_{5}$ and $\bar{\varphi}$ have vanished while counter terms for them remain. These will be needed since one-loop corrections do generate cross-terms. The dimensionally reduced version of the tree-level quadratic terms are
\begin{gather}
\sum_{n=-\infty}^{\infty}
\Bigg[
\frac{1}{2}
A_{\mu}^{(n)}
[ \square_{4} + m_{A}^{2} + n^{2} M^{2}]
A^{\mu (-n)}
-
\frac{1}{2}
A_{5}^{(n)}
\left[
\square_{4}
+
m_{A}^{2}
+
n^{2} M^{2}
\right]
A_{5}^{(-n)}
\nn \\
-
\frac{1}{2}
\varphi^{(n)}
[
\square_{4} 
+
m_{A}^{2}
+
n^{2} M^{2}
]
\varphi^{(-n)}
\Bigg].
\label{GaugeFixAction}
\end{gather}
After dimensional reduction on $S^{1}$ Lorentz invariance is broken and therefore there is no guarantee that the field redefinitions of the 4-d gauge field and the $A_{5}$ are equal. The dimensionally reduced $A_{\mu}$, $A_{5}$ and $\varphi$ quadratic counter-terms are 
\begin{gather}
\frac{1}{2}
A_{\mu}
\left[
g^{\mu\nu}
[ \delta_{A} \square_{4} + m_{A}^{2}( 2 \delta_{e} + \delta_{\phi} + \delta_{A} )]
-
\delta_{A}
\partial^{\mu}
\partial^{\nu}
\right]
A_{\nu}
-
m_{A}( \delta_{e} + \delta_{\phi} + \frac{1}{2} \delta_{A} )
A_{\mu}
\partial^{\mu} 
\varphi
\nn \\
-
\frac{1}{2}
A_{5}
\left[
\delta_{5}
\square_{4}
+
m_{A}^{2}
(2 \delta_{e} + \delta_{\phi} + \delta_{5} )
\right]
A_{5}
-
\frac{1}{2}
\varphi
[
\delta_{\phi} \square_{4} 
+
\frac{\delta \sigma}{v}
]
\varphi.
\end{gather}
Again, we have only given the counter terms for the zero mode fields since they are the only fields that appear in the LPEA. Note that the although the $\varphi$ zero mode has a nonzero mass it receives no additional mass counter terms. This fact is due to BRST symmetry, which protects the $\varphi$ from receiving non-BRST invariant corrections.

We still have to worry about ghost fields, which unfortunately to not decouple from the rest of the action in this gauge choice. Our gauge fixing condition is $G = 0$ where $G$ is  
\begin{gather}
G= 
\partial_{\mu} \bar{A}^{\mu} + \partial_{5} \bar{A}^{5} + m_{A} \bar{\varphi}.
\end{gather}
In order to find the Faddeev-Popov ghost action we need to find the functional derivative of $G$ with respect to the gauge choice. Under an infinitesimal gauge transformation:
\begin{gather}
\delta \bar{\varphi}
=
-\Lambda (\bar{\chi} + \bar{v})
,
\quad
\delta \bar{A}^{\mu}
=
- \frac{1}{\bar{e}}
\partial^{\mu} \Lambda
,
\quad
\delta \bar{A}_{5}
=
- \frac{1}{\bar{e}}
\partial_{5} \Lambda.
\end{gather}
$G$ changes by
\begin{gather}
\delta G = \frac{1}{\bar{e}}
(- \square_{4} + \partial_{5}^{2}  - \bar{e} m_{A} ( \bar{v} + \bar{\chi} ) )
\Lambda.
\end{gather} 
The Faddeev-Popov determinant is therefore
\begin{gather}
\det\left(
\frac{\delta G}{\delta \Lambda}
\right)
=
\det\left[- \square_{4} + \partial_{5}^{2} - m_{A}^{2} - \bar{e} m_{A} \bar{\chi}\right]
=
\int 
[\D \bar{c}^{\dagger} \D \bar{c}]
e^{i \int d^{5}x \bar{c}^{\dagger} [- \square_{4} + \partial_{5}^{2} - m_{A}^{2} - \bar{e} m_{A} \bar{\chi}] \bar{c}}
\nn \\
\Rightarrow
\quad
S_{gh}
=
-
\int d^{5}x \bar{c}^{\dagger} [ \square_{4} - \partial_{5}^{2} + m_{A}^{2} + \bar{e} m_{A} \bar{\chi}] \bar{c}.
\end{gather}
The dimensionally reduced form of this action is 
\begin{gather}
S_{gh}
=
-
\int d^{4}x
\left[
\sum_{n= - \infty}^{\infty}
c^{\dagger (n)}[\square_{4} + m_{A}^{2} + n^{2} M^{2} ]c^{(-n)}
+
e m_{A} \sum_{n,m=-\infty}^{\infty} c^{ \dagger (n)}c^{(m)} \chi^{(-n-m)}
\right].
\label{GhostAction}
\end{gather}
Note that like all the other fields in the action, the ghosts have a KK mass term allowing us to separate the ghost zero mode into the light sector, and the nonzero mode ghosts into the heavy sector. In the low-energy theory, the heavy mode ghosts are integrated out along with the other heavy fields, while the zero mode ghost is left in the LPEA. 

The complete action is therefore the sum of the original action (\ref{OriginalDRedAction}), the gauge fixing action (\ref{GaugeFixAction}) and the ghost action (\ref{GhostAction}). With the full gauge fixed action, it is a simple matter of generalizing the formula (\ref{KKmodeLPEAExpansion}) for gauge fields to find the LPEA. However, there is an additional subtly to the effective action formula (\ref{KKmodeLPEAExpansion}) when dealing with the anticommuting ghost fields. The formula for the LPEA for a gauge theory is given by
\begin{gather}
\Gamma[\Phi]
=
S[\Phi]
+
i
\sum_{n=1}^{\infty}
\Tr\log
\left[
K^{(n)}(\Phi) [K^{(n)}(0)]^{-1}
\right]
-
2 i
\sum_{n=1}^{\infty}
\Tr\log
\left[
K^{(n)}_{gh}(\Phi) [K_{gh}^{(n)}(0)]^{-1}
\right]
\label{GhostLPEA}
\end{gather}
where $\Phi$ is a compact notation for the set of fields $\{\varphi, \chi, A_{\mu}, A_{5}, c, c^{\dagger}\}$. The ghost $K$-matrix $K_{gh}$ is defined as
\begin{gather}
K_{gh}^{(n)}(\Phi)
=
\frac{\delta^{2} S_{gh}}{\delta c^{\dagger (n)} \delta c^{(-n)}}.
\end{gather}
The minus sign in front of the ghost contribution to the LPEA (\ref{GhostLPEA}) is due to the anticommuting nature of the ghost fields. Having laid out our gauge fixing prescription and finding the resulting ghost action, we are now in a position to find the one-loop LPEA.

\subsection{Heavy One-Loop Corrections}
The complete dimensionally reduced, gauge fixed action of the $5$-d Abelian Higgs model is
{\setlength{\baselineskip}{2.5\baselineskip}
\begin{gather}
S_{0}
+
S_{GF}
+
S_{gh}
=
\int d^{4}x
\left[
\sum_{n = -\infty}^{\infty}
\left(
\frac{1}{2}
A_{\mu}^{(n)}
[
\square_{4}
+
m_{A}^{2}
+
n^{2} M^{2}
]A^{\mu (-n)}
\right.
\right.
\nn \\
-
\frac{1}{2} A_{5}^{(n)}
\left[
\square_{4}
+
m_{A}^{2} 
+
n^{2} M^{2}
\right]
A_{5}^{(-n)}
-
\frac{1}{2} \varphi^{(n)} \left[ \square_{4} + m_{A}^{2} + n^{2} M^{2}\right] \varphi^{(-n)} 
\nn \\
\left.
-
\frac{1}{2} \chi^{(n)} \left[ \square_{4} + m_{\chi}^{2} + n^{2} M^{2}\right] \chi^{(-n)} 
-
c^{\dagger (n)} \left[ \square_{4} + m_{A}^{2} + n^{2} M^{2}\right] c^{(-n)} 
\right)
\nn \\
+
\sum_{k,l=-\infty}^{\infty}
\left[
e
A_{\mu}^{(k)}
\left(
\varphi^{(l)} \partial^{\mu}\chi^{(-k-l)}
-
\chi^{(l)} \partial^{\mu}\varphi^{(-k-l)}
\right)
+
i e  (k+ l) M
A_{5}^{(k)} 
\left(
\varphi^{(l)} \chi^{(-k-l)}
-
\chi^{(l)} \varphi^{(-k-l)}
\right)
\right.
\nn \\
\left.
+
e m_{A}
\chi^{(k)}  
\left(A_{\mu}^{(l)} A^{\mu (-k-l)}
-
A_{5}^{(l)} A_{5}^{(-k-l)}
\right)
-
\frac{\eta}{3!}
(\varphi^{(k)} \varphi^{(l)} + \chi^{(k)} \chi^{(l)})\chi^{(-k-l)}
-
e m_{A} c^{\dagger (k)}
c^{(l)}
\chi^{(-k-l)}
\right]
\nn \\
+
\sum_{k,l,n=-\infty}^{\infty}
\left(
\frac{e^{2}}{2}
\left(
\varphi^{(k)} \varphi^{(l)} 
+
\chi^{(k)} \chi^{(l)} 
\right)
\left(
A_{\mu}^{(n)} A^{\mu (-k-l-n)}
-
A_{5}^{(n)} A_{5}^{(-k-l-n)}
\right)
\right.
\nn \\
\left.
\left.
-
\frac{\lambda}{4!}
\left(\varphi^{(k)} \varphi^{(l)} \varphi^{(n)} \varphi^{(-k-l-n)} + 2 \chi^{(k)} \chi^{(l)} \varphi^{(n)} \varphi^{(-k-l-n)} + \chi^{(k)} \chi^{(l)} \chi^{(n)} \chi^{(-k-l-n)} \right)
\right)
\right].
\end{gather}}
The counter term action is 
{\setlength{\baselineskip}{2.5\baselineskip}
\begin{gather}
\delta S
=
\int d^{4}x
\Bigg[
\frac{1}{2}
A_{\mu}
\Big(
g^{\mu\nu}
[\delta_{A} \square_{4}  + m_{A}^{2} (2 \delta_{e} + \delta_{v} + \delta_{\phi} + \delta_{A})]
-
\delta_{A} \partial^{\mu}
\partial^{\nu}
\Big)
A_{\nu}
\nn \\
-
\frac{1}{2}
A_{5}
[\delta_{5} \square_{4} + m_{A}^{2}(2 \delta_{e} + \delta_{v} + \delta_{\phi} + \delta_{5})]
A_{5}
-
m_{A} \left( \delta_{e} + \frac{1}{2} \delta_{v} + \delta_{\phi} +\frac{1}{2} \delta_{A} \right)
A_{\mu}  \partial^{\mu} \varphi
\nn \\
-
\frac{1}{2} \varphi  \left[ \delta_{\phi} \square_{4} + \frac{\delta \sigma}{v}\right] \varphi 
-
\frac{1}{2}  \chi \left[ \delta_{\phi} \square_{4} + m_{\chi}^{2}( \delta_{\lambda}  + \delta_{v} + 2 \delta_{\phi}) + \frac{\delta \sigma}{v} \right] \chi
\nn \\
-
\delta \sigma \chi
+
e 
\left(
\delta_{e}
+
\delta_{\phi}
+
\frac{1}{2} \delta_{A}
\right)
A_{\mu}
\left(
\varphi \partial^{\mu}\chi
-
\chi \partial^{\mu}\varphi
\right)
\nn \\
+
e m_{A} \left(2 \delta_{e} + \frac{1}{2}\delta_{v} + \delta_{\phi} + \delta_{A}\right)
\chi
\left(
A_{\mu} A^{\mu}
-
A_{5}^{2}
\right)
-
\frac{\eta}{3!}
\left( \delta_{\lambda} + \frac{1}{2}\delta_{v} + 2 \delta_{\phi}\right)
\chi
(\varphi^{2} + \chi^{2})
\nn \\
+
\frac{e^{2}}{2}
\left(2 \delta_{e}
+
\delta_{\phi}
+
\delta_{A}
\right)
\left(
\varphi^{2} 
+
\chi^{2} 
\right)
\left(
A_{\mu} A^{\mu}
-
A_{5}^{2}
\right)
-
\frac{\lambda}{4!}
\left(\delta_{\lambda}+ 2 \delta_{\phi}\right)
(\varphi^{2} + \chi^{2} )^{2}
\Bigg].
\label{COUNTERTERMACTION}
\end{gather}}
After a tedious calculation integrating out the heavy modes using the program described in appendix \ref{LPEAphi4Theory}, we find the one-loop LPEA for the zero mode sector: 
\begin{gather}
\int d^{4}x
\left(
-
\frac{\Gamma_{A}}{4}
F_{\mu\nu}
F^{\mu\nu}
+
\frac{\Gamma_{m_{A}}^{(A_{\mu})} m_{A}^{2} }{2}
A_{\mu} A^{\mu}
-
\frac{1}{2} A_{5}
\left[
\Gamma_{5}
\square_{4}
+
\Gamma_{m_{A}}^{(A_{5})} m_{A}^{2}
\right]
A_{5}
\right.
\nn \\
+
\frac{\Gamma_{\phi}}{2} \partial_{\mu} \varphi \partial^{\mu}\varphi
-
\frac{\Gamma_{\varphi^{2}}}{2}
\varphi^{2}
+
\frac{\Gamma_{\phi}}{2} \partial_{\mu} \chi \partial^{\mu}\chi
-
\frac{\Gamma_{m_{\chi}} m_{\chi}^{2}}{2} \chi^{2}
-
\sigma \chi
-
\Gamma_{\varphi A}
m_{A} A_{\mu} \partial^{\mu} \varphi
\nn \\
+
\Gamma_{e}
e
A_{\mu}
(\varphi \partial^{\mu}\chi
-
\chi \partial^{\mu}\varphi)
+
\Gamma_{e m _{A}}^{(A_{\mu})}
e m_{A}
\chi
A_{\mu} A^{\mu}
-
\Gamma_{e m _{A}}^{(A_{5})}
e m_{A}
\chi
A_{5}^{2}
-
\frac{\Gamma_{\eta} \eta}{3!}
\chi (\varphi^{2} + \chi^{2})
\nn \\
\left.
+
\frac{\Gamma_{e^{2}}^{(A_{\mu})}
e^{2}}{2}
(\varphi^{2} + \chi^{2})
A_{\mu} A^{\mu}
-
\frac{\Gamma_{e^{2}}^{(A_{5})}
e^{2}}{2}
(\varphi^{2} + \chi^{2})
A_{5}^{2}
-
\Gamma_{A_{5}^{4}}
A_{5}^{4}
-
\frac{\Gamma_{\lambda} \lambda}{4!}
(\varphi^{2} + \chi^{2})^{2}
\right)
\end{gather}
where the infinite contributions are\footnote{In this paper we have used $\frac{1}{\epsilon} \left(\frac{\mu}{M}\right)^{\epsilon}$ as a short hand for $\frac{1}{\epsilon} - \log \left[\frac{M}{\mu}\right]$. The correspondence is not exact, but it is acceptable since we are only concerned with the divergent and log parts of the corrections.}:
{\setlength{\baselineskip}{2.5\baselineskip}
\begin{gather}
\Gamma_{A} 
=
-
\frac{e^{2}}{24 \pi^{2} \epsilon}
\left(
\frac{\mu}{M}
\right)^{\epsilon}
,
\qquad 
\Gamma_{m_{A}}^{(A_{\mu})}
=
\frac{e^{2}}{2 \pi^{2} \epsilon}
\left(
\frac{\mu}{M}
\right)^{\epsilon}
,
\qquad
\Gamma_{5}
=0
,
\nn \\
\Gamma_{m_{A}}^{(A_{5})}m_{A}^{2}
=
\frac{e^{2} (5 m_{A}^{2} + m_{\chi}^{2})}{8 \pi^{2} \epsilon}
\left(
\frac{\mu}{M}
\right)^{\epsilon}
+
\frac{3 e^{2} M^{2} \zeta(3)}{8 \pi^{4}}
,
\qquad
\Gamma_{\phi}
=
\frac{e^{2}}{4 \pi^{2} \epsilon}
\left(
\frac{\mu}{M}
\right)^{\epsilon}
,
\nn \\
\Gamma_{\varphi^{2}}
=
\frac{3 \lambda m_{\chi}^{2} + \lambda m_{A}^{2} + 24 e^{2} m_{A}^{2}}{48 \pi^{2} \epsilon}
\left(
\frac{\mu}{M}
\right)^{\epsilon}
+
\frac{(\lambda + 6 e^{2}) M^{2} \zeta(3)}{24 \pi^{4}},
\nn \\
\Gamma_{m_{\chi}}m_{\chi}^{2}
=
\frac{13 \lambda m_{\chi}^{2} + \lambda m_{A}^{2} + 72 e^{2} m_{A}^{2}}{48 \pi^{2} \epsilon}
\left(
\frac{\mu}{M}
\right)^{\epsilon}
+
\frac{(\lambda + 6 e^{2}) M^{2} \zeta(3)}{24 \pi^{4}},
\nn \\
\frac{\sigma}{v}
=
\frac{3 \lambda m_{\chi}^{2} + \lambda m_{A}^{2} + 24 e^{2} m_{A}^{2}}{48 \pi^{2} \epsilon}
\left(
\frac{\mu}{M}
\right)^{\epsilon}
+
\frac{(\lambda + 6 e^{2}) M^{2} \zeta(3)}{24 \pi^{4}},
\nn \\
\Gamma_{\varphi A}
=
\frac{3 e^{2}}{8 \pi^{2} \epsilon}\left(
\frac{\mu}{M}
\right)^{\epsilon}
,
\qquad
\Gamma_{e}
=
\frac{e^{2}}{4 \pi^{2} \epsilon}
\left(
\frac{\mu}{M}
\right)^{\epsilon}
,
\qquad
\Gamma_{em_{A}}^{(A_{\mu})}
=
\frac{3 e^{2}}{8 \pi^{2} \epsilon}
\left(
\frac{\mu}{M}
\right)^{\epsilon},
\nn \\
\Gamma_{em_{A}}^{(A_{5})}
=
\frac{\lambda + 6 e^{2}}{12 \pi^{2} \epsilon}
\left(
\frac{\mu}{M}
\right)^{\epsilon}
,
\qquad
\Gamma_{\eta} \eta
=
\frac{5 \eta \lambda - 3 e^{2} \eta + 72 e^{3} m_{A}}{24 \pi^{2} \epsilon}
\left(
\frac{\mu}{M}
\right)^{\epsilon}
,
\nn \\
\Gamma_{e^{2}}^{(A_{\mu})}
=
\frac{e^{2}}{4 \pi^{2} \epsilon}
\left(
\frac{\mu}{M}
\right)^{\epsilon}
,
\quad
\Gamma_{e^{2}}^{(A_{5})}
=
\frac{\lambda + 6 e^{2}}{12 \pi^{2} \epsilon}
\left(
\frac{\mu}{M}
\right)^{\epsilon}
,
\nn \\
\Gamma_{A_{5}^{4}}
=
\frac{e^{4}}{16 \pi^{2} \epsilon}
\left(
\frac{\mu}{M}
\right)^{\epsilon}
,
\qquad
\Gamma_{\lambda}  \lambda
=
\frac{5 \lambda^{2} - 6  \lambda e^{2} + 72 e^{4}}{24 \pi^{2} \epsilon}
\left(
\frac{\mu}{M}
\right)^{\epsilon}.
\label{ZCorrections}
\end{gather}}
It should be noted that the $M^{2}$ piece in the $A_{5}$ mass correction $\Gamma_{m_{A}}^{(A_{5})}$ has been calculated previously, though in a different context \cite{Cheng:2002iz, Puchwein:2003jq}. From this result, it is immediately clear that the corrections to the $|\phi|^{2} A_{\mu} A^{\mu}$ and $|\phi|^{2} A_{5}^{2}$ couplings are different, which implies that the charge receives a different correction at different vertices. If we were to find the correction to the electric charge by evaluating corrections to the $|\phi|^{2} A_{5}^{2}$ vertex we would find that the divergent one-loop correction to the charge is
\begin{gather}
\delta e^{2}
=
\frac{e^{2} (\lambda + 3 e^{2})}{12 \pi^{2} \epsilon}
\left(
\frac{\mu}{M}
\right)^{\epsilon}
.
\label{chargeA5Correction}
\end{gather}
This is in contrast to the charge correction that is obtained from the $|\phi|^{2} A_{\mu} A^{\mu}$ vertex:
\begin{gather}
\delta e^{2}
=
\frac{e^{4}}{24 \pi^{2} \epsilon}
\left(
\frac{\mu}{M}
\right)^{\epsilon}.
\end{gather}
By now this should come as no surprise. The same corrections were found for the same model in the symmetric phase \cite{Akhoury:2007dz}. The correction to the charge in (\ref{chargeA5Correction}) signals a breakdown of charge universality since there is a dependence of the result on $\lambda$, which in turn depends on the matter field $\phi$. 

The reason for the differing charge corrections is the absence of a gauge symmetry protecting $A_{5}$. With no Ward identities, there is no guarantee that $\Gamma_{e^{2}}^{(A_{5})}$ be related to $\Gamma_{\phi}$. In contrast, the local operators involving $A_{\mu}$ do satisfy 4-d Ward identities to one-loop order, leading to the equalities: $\Gamma_{\phi} = \Gamma_{e}$ and $\Gamma_{e^{2}}^{(A_{\mu})} = \Gamma_{e}$, which are indeed satisfied by the corrections found in (\ref{ZCorrections}). Thus we have no right to expect that the correction to $e^{2}$ at the $|\phi|^{2} A_{5}^{2}$ vertex will be the same as at the $|\phi|^{2} A_{\mu} A^{\mu}$ vertex.

\section{Subtraction of the Mass and Coupling Divergences}
\label{MassandCouplingRenormalization}
The divergences of the one-loop corrections are eliminated if the following hold:
\begin{gather}
\sigma + \delta \sigma
=0,
\\
\Gamma_{\phi} + \delta_{\phi} 
=0,
\\
\Gamma_{\varphi^{2}}
+
\frac{\delta \sigma}{v}
=0,
\\
\Gamma_{m_{\chi}} m_{\chi}^{2}
+
(\delta_{\lambda} + \delta_{v} + 2 \delta_{\phi})
m_{\chi}^{2}
+
\frac{\delta \sigma}{v}
=0,
\\
\Gamma_{A}
+
\delta_{A}
=0,
\\
\Gamma_{m_{A}}^{(A_{\mu})}
+
\left(
2 \delta_{e}
+
\delta_{v}
+
\delta_{\phi}
+
\delta_{A}
\right)
=0,
\\
\Gamma_{5} + \delta_{5} = 0
\\
\Gamma_{m_{A}}^{(A_{5})}
+
\left(
2 \delta_{e}
+
\delta_{v}
+
\delta_{\phi}
+
\delta_{5}
\right)
=0,
\label{A5massRenorm}
\\
\Gamma_{\varphi A}
+
(\delta_{e} + \frac{1}{2} \delta_{v} + \delta_{\phi} + \frac{1}{2} \delta_{A})
=0,
\\
\Gamma_{e}
+
\left(
\delta_{e} 
+
\delta_{\phi}
+
\frac{1}{2} \delta_{A}
\right)
=0,
\\
\Gamma_{e m_{A}}^{(A_{\mu})}
+
\left(
2 
\delta_{e}
+
\frac{1}{2}
\delta_{v}
+
\delta_{\phi}
+
\delta_{A}
\right)
=0,
\\
\Gamma_{e m_{A}}^{(A_{5})}
+
\left(
2 
\delta_{e}
+
\frac{1}{2}
\delta_{v}
+
\delta_{\phi}
+
\delta_{5}
\right)
=0,
\label{A5chiRenorm}
\\
\Gamma_{\eta}
+
\left(
\delta_{\lambda}
+
\frac{1}{2}\delta_{v}
+
2
\delta_{\phi}
\right)=0,
\\
\Gamma_{e^{2}}^{(A_{\mu})}
+
\left(
2
\delta_{e}
+
\delta_{\phi}
+
\delta_{A}
\right)
=0,
\\
\Gamma_{e^{2}}^{(A_{5})}
+
\left(
2
\delta_{e}
+
\delta_{\phi}
+
\delta_{5}
\right)
=0,
\label{A5chichiRenorm}
\\
\Gamma_{\lambda}
+
\left(
\delta_{\lambda}
+
2
\delta_{\phi}
\right)
=0.
\end{gather}
Please note that there is no choice of counter terms that allow for all these equations to be satisfied simultaneously. As was found in \cite{Akhoury:2007dz}, the $A_{5}$ is the source of the impediment to consistently subtracting divergence from the one-loop corrections. If we ignore the $A_{5}$ vertex corrections then all the remaining divergences can be subtracted if we choose\footnote{Our results for the counter terms should be compared with those found in \cite{Huang:1991wn}}:
\begin{gather}
\delta \sigma
=
-
\frac{v^{3}
\left(
\lambda^{2} + \lambda e^{2} + 24 e^{4} 
\right)}{48 \pi^{2} \epsilon}
\left(
\frac{\mu}{M}
\right)^{\epsilon}
-
\frac{v
\left(
\lambda + 6 e^{2}
\right) M^{2} \zeta(3)}{24 \pi^{4}}
,
\label{CT1}
\\
\delta_{v}
=
-
\frac{e^{2}}{4 \pi^{2} \epsilon}
\left(
\frac{\mu}{M}
\right)^{\epsilon}
,
\quad
\delta_{\phi}
=
-
\frac{e^{2}}{4 \pi^{2} \epsilon}
\left(
\frac{\mu}{M}
\right)^{\epsilon}
,
\qquad
\delta_{A}
=
\frac{e^{2}}{24 \pi^{2} \epsilon}
\left(
\frac{\mu}{M}
\right)^{\epsilon}
,
\label{CT2}
\\
\delta_{e}
=
-
\frac{e^{2}}{48 \pi^{2} \epsilon}
\left(
\frac{\mu}{M}
\right)^{\epsilon}
,
\qquad
\lambda \delta_{\lambda}
=
-
\frac{5 \lambda^{2} - 18 \lambda e^{2} + 72 e^{4}}{24 \pi^{2} \epsilon}
\left(
\frac{\mu}{M}
\right)^{\epsilon}.
\label{CT3}
\end{gather}
Here we have chosen to work in a modified minimal subtraction scheme. We have subtracted the $\frac{1}{\epsilon}$ pole from each of the divergences, but we have also subtracted the finite $\log M/\mu$ piece. Finite constants like $\gamma$ and  $\log 4 \pi$ have also been subtracted, but their presence is unimportant in our results. Indeed, in order to show the decoupling we must go beyond the MS scheme.

We should compare the counter terms to those found in the 5-d extension of scalar QED \cite{Akhoury:2007dz}. One can see immediately that the counter terms for the couplings: $e$ and $\lambda$, and the field redefinitions are the same for the symmetric and broken phases. Before we can compare the mass counter terms we need to relate $\delta \sigma$ and $\delta_{\lambda}$ to the mass squared counter term $\delta_{m^{2}}$ from the symmetric phase. Recall that 
\begin{gather}
\delta \sigma = \frac{\lambda v^{3}}{6} (\delta_{v} + \delta_{\phi} + \delta_{\lambda}  - \delta_{m^{2}}).
\label{SymBrokCTRelationOne}
\end{gather}
Symanzik's theorem \cite{Symanzik:1969ek,Symanzik:1970} implies that this relation should hold true up to the divergent parts of the counter terms. The counter term used to subtract the divergences from the mass correction in the symmetric phase is
\begin{gather}
m^{2}
\delta_{m^{2}} 
=
\frac{m^{2} (9 e^{2} - 2 \lambda)}{24 \pi^{2} \epsilon}
\left(
\frac{\mu}{M}
\right)^{\epsilon}
-
\frac{(\lambda + 6 e^{2}) M^{2} \zeta(3)}{24 \pi^{4}}
\label{mSquaredCT}
\end{gather}
Using the counters terms (\ref{CT1})-(\ref{CT2}), we find that the relation (\ref{SymBrokCTRelationOne}) is indeed satisfied up to the divergent corrections. Ignoring the $A_{5}$ sector for the moment, the counter terms used here are the same as those used to subtract the divergences from the LPEA in the symmetric phase \cite{Akhoury:2007dz}. Since the divergences in the symmetric phase can be consistently subtracted, the equality of the counter terms between phases implies that the divergences in the broken phase, with it's massive gauge boson, can also be consistently subtracted. Unfortunately, as we will see later, the same can not be said for the $A_{5}$ sector, which has markedly different divergences in the broken and symmetric phases.

\begin{figure}
\begin{center}
\includegraphics[scale=.75]{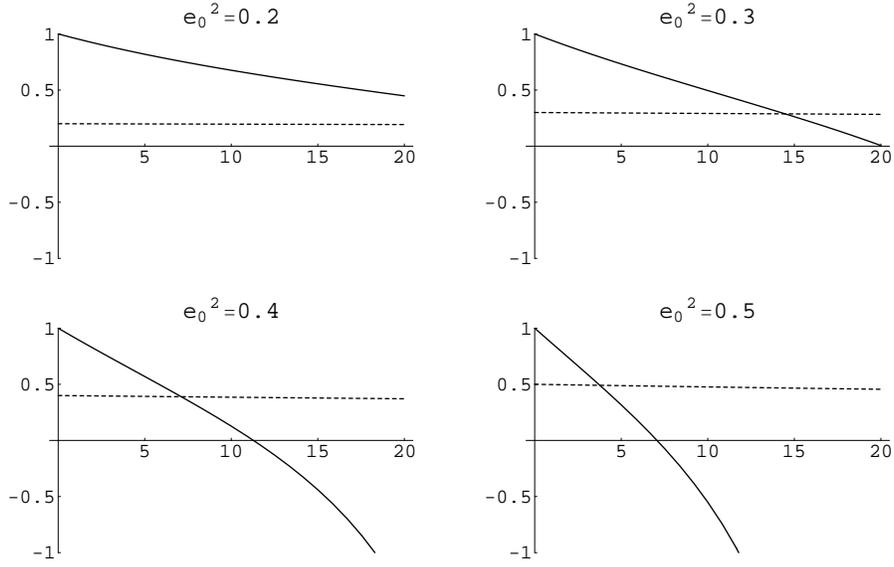}
\caption{These four plots show the $M$-scaling behavior of the scalar coupling $\lambda$ (solid line) and the gauge coupling squared $e^{2}$ (dashed line) versus $\log_{10} M/M_{0}$. Each plot shows the running of both couplings for different values of the gauge coupling at $M = M_{0}$.}
\label{LambdaChargeVersusLogM}
\end{center}
\end{figure}

An interesting question arises about the loop correction's dependence on the compactification scale $M$. Consider the bare couplings to one-loop order:
\begin{gather}
e^{2}_{b} 
=
e^{2} (1 + 2 \delta_{e}) 
=
e^{2} -  \frac{e^{4}}{24 \pi^{2}\epsilon} 
\left(
\frac{\mu}{M}
\right)^{\epsilon},
\\
\lambda_{b} 
=
\lambda (1 + \delta_{\lambda}) 
= 
\lambda
-
\frac{5 \lambda^{2} - 18 \lambda e^{2} + 72 e^{4}}{24 \pi^{2} \epsilon}
\left(
\frac{\mu}{M}
\right)^{\epsilon}
.
\end{gather} 
Since the bare couplings are intrinsic parameters of the theory, they are independent of the scale $M$. Therefore, we can differentiate the relations above with respect to $\log M$ and construct a set of equations for the running of the effective couplings:
\begin{gather}
\frac{d e^{2}}{d \log M} = - \frac{e^{4}}{24 \pi^{2}},
\label{eRGESol}
\\
\frac{d \lambda}{d \log M} 
=  
-\frac{5 \lambda^{2} - 18 \lambda e^{2} + 72 e^{4}}{24 \pi^{2}}.
\label{LambdaRGESol}
\end{gather}
It is interesting to note that the coefficient of $e^{4}$ in (\ref{LambdaRGESol}) is in general equal to $18 (d-1)$, where $d$ is the {\it total} number of space-time dimensions. Even though this is an effective theory in four dimensions the couplings still ``feel'' the effects of the 5-th dimension. Without solving for the couplings, we can already see that they become weaker as the scale $M$ increases. This is obvious for the charge coupling, and it is also true for $\lambda$ since $5 \lambda^{2} - 18 \lambda e^{2} + 72 e^{4} >0$ for all values of $e$ and $\lambda$. The solution for these coupled equations is complicated, but an analytic result does exist:
\begin{gather}
e^{2}(M)
=
\frac{e_{0}^{2}}{1 +\frac{e_{0}^{2}}{24 \pi^{2}} 
\log\left(
\frac{M}{M_{0}} \right) },
\\
\lambda(M)
=
e^{2}(M)
\left(
\frac{19}{10} - \frac{\sqrt{1079}}{10}
\tan\left[
z(M,M_{0},\lambda_{0},e_{0})
\right]
\right)
\end{gather}
where
\begin{gather}
z(M,M_{0},\lambda_{0},e_{0})
=
\textrm{arctan}
\left[
\frac{1}{\sqrt{1079}}
( 19 - \frac{10 \lambda_{0}}{e_{0}^{2}})
\right]
+
\frac{\sqrt{1079}}{2}
\log\left[
\frac{e_{0}^{2}}{e^{2}(M)}
\right].
\end{gather}
The scaling behavior of the couplings is shown in figure \ref{LambdaChargeVersusLogM}. As the energy scale increases the two couplings become weaker, though $\lambda$ decreases at a much faster rate than $e^{2}$. As a result, if at a low scale $\lambda_{0} > e^{2}_{0}$, they will eventually intersect at some high scale. As $M$ continues to increase, the scalar coupling becomes negative, leading to an unstable system, and then diverges when $z = \frac{\pi}{2}$. This is acceptable since our analysis ignores gravity, and thus the LPEA is not expected to be a valid description of the high-energy physics.

\begin{figure}
\begin{center}
\includegraphics[scale=.75]{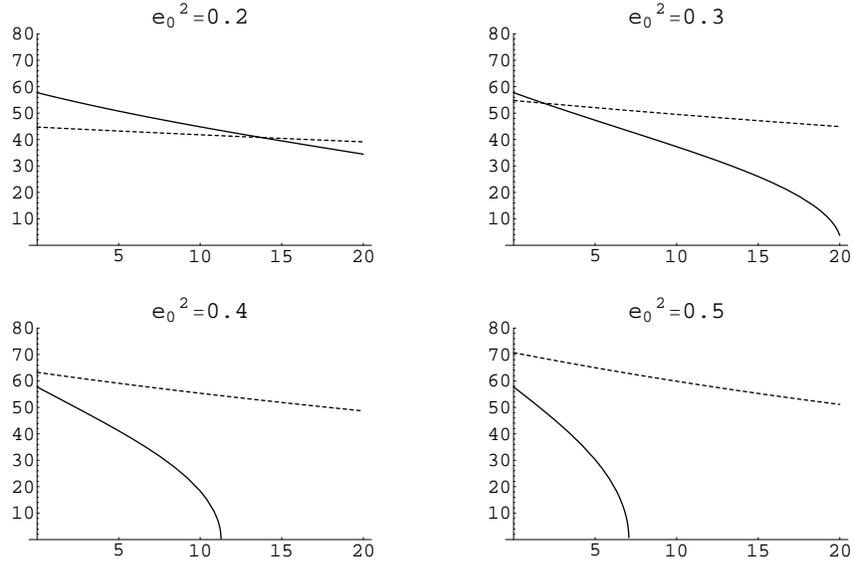}
\caption{These four plots show the $M$-scaling behavior of the Higgs mass $m_{\chi}^{2}$ (solid line) and the gauge mass $m_{A}$ (dashed line) versus $\log_{10} M/M_{0}$. Each plot shows the running of both masses for different values of the gauge coupling at $M = M_{0}$.}
\label{HiggsGaugeMassVersusLogM}
\end{center}
\end{figure}

The $M$-scale running equations for the effective Higgs and gauge masses can be similarly determined. Recall that the Higgs and gauge masses are the product of the VEV $v$, and the scalar and gauge couplings, respectively. Therefore the mass runnings are related to the coupling and VEV runnings by:
\begin{gather}
\frac{d m_{A}^{2}}{d \log M}
=
v^{2}
\frac{d e^{2}}{d \log M}
+
2 e^{2} v
\frac{d v}{d \log M}
,
\\
\frac{d m_{\chi}^{2}}{d \log M}
=
\frac{v^{2}}{3}
\frac{d \lambda}{d \log M}
+
\frac{2 \lambda v}{3}
\frac{d v}{d \log M}.
\end{gather}
The gauge and scalar coupling beta functions have already been obtained in (\ref{eRGESol}) and (\ref{LambdaRGESol}). The running of the VEV can be found by appropriately generalizing the result in \cite{PhysRevD.10.2530}:
\begin{gather}
\frac{d \log v}{d \log M}
=
-
\frac{1}{2} \gamma_{\phi}(\lambda,e).
\end{gather}
Here $\gamma_{\phi}$ is the anomalous dimension of the scalar field with respect to changes in the compactification scale. The anomalous $M$-scaling dimension is defined in terms of the $\phi$ field redefinition as
\begin{gather}
\gamma_{\phi}
=
\lim_{\epsilon \rightarrow 0}
\frac{d \log Z_{\phi}}{d \log M}.
\end{gather}
Therefore, the running of the VEV to one-loop order is
\begin{gather}
\frac{d \log v}{d \log M}
=
-
\frac{e^{2}}{8 \pi^{2}}.
\end{gather}
It follows that the running of the effective masses are,
\begin{gather}
\frac{d m_{A}^{2}}{d \log M}
=
-
\frac{7 e^{2} m_{A}^{2}}{24 \pi^{2}}
,
\\
\frac{d m_{\chi}^{2}}{d \log M}
=
-\frac{5 \lambda m_{\chi}^{2} - 12 e^{2} m_{\chi}^{2} + 24 e^{2} m_{A}^{2}}{24 \pi^{2}}.
\end{gather}
The solutions are just the product of the appropriate coupling with the VEV:
\begin{gather}
m_{A}^{2}(M)
=
e^{2}(M) v^{2}(M),
\\
m_{\chi}^{2}(M)
=
\frac{\lambda(M) v^{2}(M)}{3}
\end{gather}
where
\begin{gather}
v(M)
=
\frac{v_{0}}{(1 + \frac{e^{2}_{0}}{24 \pi^{2}} \log M /M_{0})^{3}}.
\end{gather}
A plot of the masses has been included in figure \ref{HiggsGaugeMassVersusLogM}. Qualitatively, the scaling behavior of the Higgs and gauge mass is the same as the scalar and gauge couplings, respectively. The only effect that the VEV's $M$-scale dependence has is to hasten each mass's decrease. Like their respective couplings, the Higgs mass decreases at a higher rate than the gauge mass. If this were the SM, we would have to fix the trajectories so that $m_{\chi} > m_{A}$ below the electroweak scale. However, no matter the initial conditions, the two masses will eventually meet at a high scale, after which the gauge mass becomes the larger of the two. If the scale continues to increase, the Higgs mass squared becomes negative and the effective theory description breaks down.

With many physically unacceptable possibilities arising in this model, it is possible that constraints may be placed on the compactification scale $M$. The fact that $m_{\chi}$ decreases at low scales and $m_{A} > m_{\chi}$ at high scales, places an upper bound on $M$, and therefore a lower bound on the compact dimension size. The physically unappealing region where $\lambda < 0$ can also be used to place an upper limit on $M$. 

However, before we can determine these constraints on $M$ we first need to understand what the scale $M_{0}$ is and how it determines the initial conditions in the trajectories. Since these theories display a type of asymptotically freedom with respect to $M$, then there is a scale $\Lambda$ analogous to $\Lambda_{QCD}$ in QCD where the coupling becomes of order 1. In the simple $\phi^{4}$ model we can write the solution for $\lambda$ as
\begin{gather}
\lambda \propto \frac{1}{\log M/\Lambda}
\end{gather}
thereby eliminating the need for an exact initial condition on $\lambda$. This will probably be possible in the more complicated Abelian Higgs model, but that has yet to be determined. In the future we will want to better understand the parameter space constraints on the compactification scale.

\subsection{Subtraction of Divergences in the $A_{5}$ Sector}

As we will define it, an LPEA finite model is one which contains all terms needed to absorb the divergent loop corrections. By this definition, our model is not LPEA finite since loop corrections have created divergences in the local operators of $A_{5}$ that can not be eliminated using the counter terms (\ref{CT1})-(\ref{CT2}). The appearance of these news divergences in the low-energy effective action should come as no surprise. Looking at (\ref{A5Transform}) it is clear that there is no gauge symmetry in four dimensions that acts on $A_{5}$. Since $A_{5}$ is a scalar with respect to the action of the 4-d Poincar\'e group, there is no reason to expect that it will not develop different mass and quartic coupling corrections than the 4-d gauge field.

In order to render the theory completely finite we have to introduce new counter terms for the $A_{5}$ vertex functions. Introducing these new counter terms will do violence to the original 5-d gauge and Lorentz invariance since they require us to separate $\bar{A}_{5}$ from the rest of the components of the gauge field, destroying covariance. If this unappealing feature is ignored, then by adding the counter terms:
\begin{gather}
-
\frac{\delta_{m_{5}^{2}}}{2} A_{5}^{2}
,
\qquad
-
\delta_{em_{A}}^{(A_{5})}
e m_{A} \chi A_{5}^{2}
,
\qquad
-\delta_{e^{2}}^{(A_{5})} e^{2}  |\phi|^{2} A_{5}^{2}
,
\qquad
-
\delta_{A_{5}^{4}}
A_{5}^{4}
\end{gather}
the theory can be made finite. For now we will ignore issues of 5-d gauge and Lorentz invariance and accept that these must be violated in order to make the theory finite (Lorentz violating counter terms could be sourced by D-brane localized interactions). Once we include the new counter terms the conditions for finiteness (\ref{A5massRenorm}), (\ref{A5chiRenorm}) and (\ref{A5chichiRenorm}) become:
\begin{gather}
\Gamma_{m_{A}}^{(A_{5})} m_{A}^{2}
+
\left(
2 \delta_{e}
+
\delta_{v}
+
\delta_{\phi}
+
\delta_{5}
\right)
m_{A}^{2}
+
\delta_{m_{5}^{2}}
=0,
\label{A5massRenormNew}
\\
\Gamma_{e m_{A}}^{(A_{5})}
+
\left(
2 
\delta_{e}
+
\frac{1}{2}
\delta_{v}
+
\delta_{\phi}
+
\delta_{5}
\right)
+
\delta_{e m_{A}}^{(A_{5})}
=0,
\label{A5chiRenormNew}
\\
\Gamma_{e^{2}}^{(A_{5})}
+
\left(
2
\delta_{e}
+
\delta_{\phi}
+
\delta_{5}
\right)
+
\delta_{e^{2}}^{(A_{5})}
=0.
\label{A5chichiRenormNew}
\end{gather}
Additionally, one-loop corrections have generated a divergent $A_{5}^{4}$ term that also needs to be subtracted by a new counter term that does not respect 5-d gauge invariance. We will label this new counter term by $\delta_{A_{5}^{4}}$. The finiteness condition on the $A_{5}^{4}$ vertex is
\begin{gather}
\Gamma_{A_{5}^{4}}
+
\delta_{A_{5}^{4}}
=0.
\label{A5tothe4Renorm}
\end{gather}
Solving (\ref{A5massRenormNew})-(\ref{A5tothe4Renorm}) for the new counter terms we find that:
\begin{gather}
\delta_{m_{5}^{2}} 
= 
\frac{e^{2} ( 3 m^{2} - m_{A}^{2})}{12 \pi^{2} \epsilon}
\left(
\frac{\mu}{M}
\right)^{\epsilon}
-
\frac{3 e^{2} M^{2} \zeta(3)}{8 \pi^{4}},
\label{A5massCT}
\\
\delta_{e m_{A}}^{(A_{5})} 
= 
-
\frac{\lambda + e^{2}}{12 \pi^{2} \epsilon}
\left(
\frac{\mu}{M}
\right)^{\epsilon},
\label{A5chiCT}
\\
\delta_{e^{2}}^{(A_{5})} 
= 
-
\frac{2 \lambda + 5 e^{2}}{24 \pi^{2} \epsilon}
\left(
\frac{\mu}{M}
\right)^{\epsilon},
\label{A5chichiCT}
\\
\delta_{A_{5}^{4}}
=
-
\frac{e^{4}}{16 \pi^{2} \epsilon}
\left(
\frac{\mu}{M}
\right)^{\epsilon}.
\label{A5tothe4CT}
\end{gather}
Clearly, the theory can still be made finite even with the $A_{5}$ field. Unfortunately, to do so we have to give the $A_{5}$ sector different counter terms than those of the $A_{\mu}$. This will have the undesirable side effect of explicitly breaking the 5-d Lorentz and gauge symmetry of the underlying theory, not including the breaking that takes place from compactification. In addition as mentioned above, we will also have problems retaining charge universality.

\subsection{Counter Terms: Broken versus Symmetric Phases}

One may wonder how the counter terms in the broken phase compare with those in the symmetric phase. From a result due to Symanzik \cite{Symanzik:1969ek,Symanzik:1970} it is expected that in most cases of spontaneous symmetry breaking, the counter terms in the broken phase are a combination of the counter terms in the symmetric phase. However, in the model under consideration, there is not an exact equivalence between the two sets of counter terms. 

The reason for this violation is again due to the $A_{5}$ zero mode. If we were to ignore the vertex operators with external $A_{5}$ legs, then the counter terms (\ref{CT1}), (\ref{CT2}), (\ref{mSquaredCT}) are the same as those found in the symmetric case \cite{Akhoury:2007dz}. The problem is therefore isolated to the $A_{5}$ sector. Putting aside issues of 5-d gauge and Lorentz invariance, the theory in the symmetric phase can be made finite by adding the counter terms:
\begin{gather}
-
\frac{\delta_{m_{5}^{2}}}{2} A_{5}^{2}
,
\qquad
-\delta_{e^{2}}^{(A_{5})} e^{2}  |\phi|^{2} A_{5}^{2}
,
\qquad
-
\delta_{A_{5}^{4}}
A_{5}^{4}.
\end{gather}
If the two sets of counter terms are equivalent, then the counter terms in the broken phase should be related to the symmetric phase counter terms like:
\begin{gather}
\delta_{A_{5}^{4}}|_{ssb}
=
\delta_{A_{5}^{4}}|_{sym},
\label{0CTrelation}
\\
\delta_{e^{2}}^{(A_{5})}|_{ssb}
=
\delta_{e^{2}}^{(A_{5})}|_{sym},
\label{1CTrelation}
\\
\delta_{e m_{A}}^{(A_{5})}|_{ssb}
=
\delta_{e^{2}}^{(A_{5})}|_{sym},
\label{2CTrelation}
\\
\delta_{m_{5}^{2}}|_{ssb}
=
\delta_{m_{5}^{2}}|_{sym}
+
m_{A}^{2}
\delta_{e^{2}}^{(A_{5})}|_{sym}.
\label{3CTrelation}
\end{gather}
However, if we calculate the symmetric phase counter terms \cite{Akhoury:2007dz}:
\begin{gather}
\delta_{m_{5}^{2}}
=
-
\frac{e^{2} m^{2}}{4 \pi^{2} \epsilon}
\left(
\frac{\mu}{M}
\right)^{\epsilon}
-
\frac{3 e^{2} M^{2} \zeta(3)}{8 \pi^{4}},
\\
\delta_{e^{2}}^{(A_{5})} 
= 
-
\frac{2 \lambda + 5 e^{2}}{24 \pi^{2} \epsilon}
\left(
\frac{\mu}{M}
\right)^{\epsilon},
\\
\delta_{A_{5}^{4}}
=
-
\frac{e^{4}}{16 \pi^{2} \epsilon}
\left(
\frac{\mu}{M}
\right)^{\epsilon}
\end{gather} 
we see that the relations (\ref{2CTrelation}) and (\ref{3CTrelation}) are not respected.

In order to understand this disconnect between the two phases we must first imagine the model as consisting of two sectors: a 4-d gauge field sector and a scalar $A_{5}$ sector. These two sectors do not interact directly, but are linked by their interaction with the Goldstone and Higgs fields. If one looks at any of these sectors individually by restricting the Higgs and Goldstone fields to interact with only one sector at the time, they would find that the divergences in the symmetric and broken phases are the same. The problem comes in when the Higgs and Goldstones interact with the two sectors at the same time.

In the symmetric phase there is no significant difference between the corrections in the 5-d Higgs model and the corrections we would find if we considered the two sectors separately. In the broken phase the Goldstone is not a true Goldstone because it is ``eaten'' by the gauge field to become the longitudinal state of the resulting massive gauge boson. In the gauge we have chosen to work in, the pseudo-Goldstone has a zero mode mass $m_{A}$. This explicitly breaks the global $U(1)$ symmetry of the $A_{5}$ sector action. Since this mass term only comes about in the broken phase, then by Symanzik's theorem there are new divergences in operators of dimension two or less. Indeed, what we find is that there is a new divergence in the $A_{5}$ mass correction when we are in the broken phase.

In general, the mass of the Goldstone is dependent upon the gauge fixing used. If we had used the Lorentz gauge the Goldstone would not have a mass but the quadratic mixing term $A_{\mu} \partial^{\mu} \varphi$ remains. In this gauge we now have to contend with mixed $\varphi-A_{\mu}$  internal lines. For example, in addition to the one-loop $\varphi$ correction to the $A_{5}$ mass: 
\begin{center}
\includegraphics{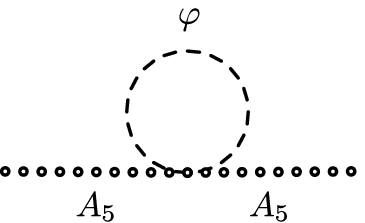}
\end{center}
we also have diagrams with mixed internal lines that also contribute:
\begin{center}
\includegraphics{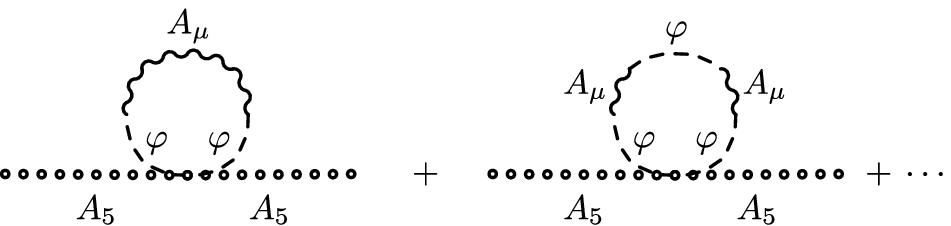}
\end{center}
Even though the Goldstone is massless in this gauge, the result of these diagrams is to give $\varphi$ an effective mass $m_{A}$. The additional divergence first found in the 't Hooft-Feynman gauge is still there in the Lorentz gauge. In some ways this is comforting. Although there are new divergences in the broken phase, the fact that they are the same in different gauges shows that the theory is not anomalous. Further checks can be made to show that the tree-level $S$-matrix elements for $\varphi - A_{5}$ scattering vanish.

\subsection{Decoupling of Heavy Modes in Orbifold Compactifications}
\label{CircleOrbifoldCpct}
A theory with SSB that has different divergences depending upon the phase that we are in is undesirable for a number of reasons. The most straight forward way to deal with this problem is to choose compactifications that do not permit an $A_{5}$ zero mode. In a $S^{1}/\mathbb{Z}_{2}$ orbifold compactifications \cite{Muck:2001yv} the boundary conditions on the components of the 5-d gauge field change to:
\begin{gather}
\bar{A}_{M}(x,y) 
=
\bar{A}_{M}(x,y+ 2 \pi n R),
\\
\bar{A}_{\mu}(x,y)
=
\bar{A}_{\mu}(x, - y),
\\
\bar{A}_{5}(x,y)
=
-
\bar{A}_{5}(x, - y).
\end{gather}
These boundary conditions lead to the Fourier series expansions of $\bar{A}_{\mu}$ and $\bar{A}_{5}$: 
\begin{gather}
\bar{A}_{\mu}(x,y)
=
\frac{A_{\mu}^{(0)}(x)}{\sqrt{2 \pi R}}
+
\sum_{n=1}^{\infty}
\frac{A^{(n)}_{\mu}(x)}{\sqrt{\pi R}}
\cos\left(n M y\right),
\\
\bar{A}_{5}(x,y)
=
\sum_{n=1}^{\infty}
\frac{A^{(n)}_{5}(x)}{\sqrt{\pi R}}
\sin\left(n M y\right).
\end{gather}
Once we integrate out the 5-th direction, the important parts of the action become:
\begin{itemize}
\item \underline{Light Sector Action}
\begin{gather}
S_{light}
=
\int d^{4}x
\Bigg(
-
\frac{1}{4}
F_{\mu\nu}
F^{\mu\nu}
+
\frac{m_{A}^{2}}{2}
A_{\mu}
A^{\mu}
- 
m_{A}
A_{\mu} \partial^{\mu} \phi
+
\frac{1}{2}
\partial_{\mu}
\varphi
\partial^{\mu}
\varphi
\nn \\
+
\frac{1}{2}
\partial_{\mu}
\chi
\partial^{\mu}
\chi
-
\frac{1}{2}
m_{\chi}^{2}
\chi^{2}
+
e m_{A} \chi A_{\mu} A^{\mu}
-
\frac{\eta}{3!} \chi (\varphi^{2} + \chi^{2})
\nn \\
+
e A_{\mu}
(
\varphi 
\partial^{\mu}
\chi
-
\chi 
\partial^{\mu}
\varphi
)
+
\frac{e^{2}}{2}
(\varphi^{2} + \chi^{2})
A_{\mu}
A^{\mu}
-
\frac{\lambda}{4!}
(\varphi^{2} + \chi^{2})^{2}
\Bigg)
\end{gather}
\item \underline{Light-Heavy Interactions}
\begin{gather}
S_{light-heavy}
=
\sum_{n = 1}^{\infty}
\int d^{4}x
\Bigg(
\frac{1}{2}
A_{\mu}^{(n)}
[
g^{\mu\nu}
[ \square_{4}
+
m_{A}^{2}
+
n^{2} M^{2}
]
-
\partial^{\mu} \partial^{\nu}
]A_{\nu}^{(n)}
\nn \\
-
\frac{1}{2} A_{5}^{(n)}
\left[
\square_{4}
+
m_{A}^{2} 
\right]
A_{5}^{(n)}
-
m_{A} 
A_{\mu}^{(n)} \partial^{\mu} \varphi^{(n)}
+
n M m_{A}
A_{5}^{(n)} \varphi^{(n)}
+
n M
A_{\mu}^{(n)} \partial^{\mu} A_{5}^{(n)} 
\nn \\
+
\frac{1}{2} \partial_{\mu} \varphi^{(n)} \partial^{\mu}\varphi^{(n)}
-
\frac{1}{2} n^{2} M^{2} \varphi^{(n) 2} 
+
\frac{1}{2} \partial_{\mu} \chi^{(n)} \partial^{\mu}\chi^{(n)}
-
\frac{1}{2} \left[m_{\chi}^{2} + n^{2} M^{2}\right] \chi^{(n) 2} 
\nn \\
+
e
\left(
A_{\mu} (\varphi^{(n)} \partial^{\mu}\chi^{(n)}
-
\chi^{(n)} \partial^{\mu}\varphi^{(n)})
+
A_{\mu}^{(n)} 
(\varphi \partial^{\mu}\chi^{(n)}
-
\chi^{(n)} \partial^{\mu}\varphi)
\right.
\nn \\
\left.
+
A_{\mu}^{(n)} 
(\varphi^{(n)} \partial^{\mu}\chi
-
\chi \partial^{\mu}\varphi^{(n)})
+
n M
A_{5}^{(n)} 
(\varphi \chi^{(n)}
-
\varphi^{(n)} \chi)
\right)
\nn \\
+
e m_{A} 
\left[
\chi
A_{\mu}^{(n)} A^{\mu (n)}
+
2 \chi^{(n)}
A_{\mu}^{(n)} A^{\mu}
-
\chi 
A_{5}^{(n) 2}
\right]
\nn \\
-
\frac{\eta}{3!}
\left[
\chi
(\varphi^{(n) 2} + \chi^{(n) 2})
+
2 \chi^{(n)}
(\varphi \varphi^{(n)} + \chi \chi^{(n)})
\right]
+
\frac{e^{2}}{2}
\left[
\left(
\varphi^{2} 
+
\chi^{2} 
\right)
A_{\mu}^{(n)} A^{\mu (n)}
\right.
\nn \\
\left.
+
4
\left(
\varphi \varphi^{(n)} 
+
\chi \chi^{(n)} 
\right)
A_{\mu}^{(n)} A^{\mu}
+
\left(
\varphi^{(n) 2} 
+
\chi^{(n) 2} 
\right)
A_{\mu} A^{\mu}
\right]
-
\frac{e^{2}}{2}
\left(
\varphi^{2} 
+
\chi^{2} 
\right)
A_{5}^{(n) 2}
\nn \\
-
\frac{\lambda}{4!}
(6 \varphi^{2} \varphi^{(n) 2} 
+ 
2 \varphi^{(n) 2} \chi^{2}
+ 
8 \varphi \varphi^{(n)} \chi \chi^{(n)}
+ 
2 \varphi^{2} \chi^{(n) 2}
+ 
6 \chi^{2} \chi^{(n) 2}  )
\Bigg).
\end{gather}
\end{itemize}
As we can see by looking at the 4-d action, there is no $A_{5}$ zero mode. The twisted boundary condition on $\bar{A}_{5}$ precludes the existence of a zero mode. Therefore, there is no additional scalar in the 4-d effective theory that will lead to different corrections for the gauge coupling $e$. All the coupling and mass corrections for the remaining zero mode fields are the same as in the $S^{1}$ compactification case, except for a factor of $\frac{1}{2}$ due to the different sums over the KK modes. Since the $A_{5}$ field was solely responsible for the appearance of new divergences, then by eliminating the $A_{5}$ zero mode we solve the problem entirely. Orbifold compactifications are already an attractive possibility since they allow for chiral fields \cite{Dienes:1998vg,Papavassiliou:2001be} and lead to realistic string models \cite{Dixon:1985jw,Dixon:1986jc}.

\section{Conclusions}
\label{Conclusions}
In this paper we have discussed the effects that heavy KK modes can have on the low-energy physics of a 5-d extension of the Abelian Higgs model. As we found in an earlier analysis of 5-d scalar QED \cite{Akhoury:2007dz} in the $\mathbb{R}^{3,1}\times S^{1}$ compactification, the heavy KK modes did not decouple in the low-energy theory due to the additional scalar $A_{5}$. In addition, it was found that there were new divergences that appear in the broken phase that were not present in the symmetric phase. The new divergences were entirely isolated to the $A_{5}$ sector, and were due to the interference of the $\varphi - A_{\mu}$ mixing on loop corrections to the $A_{5}$ mass. In the gauge we have chosen to work in the $\varphi$ field has a mass $m_{A}$. Since the pseudo-Goldstone has a mass, diagrams like
\begin{center}
\includegraphics{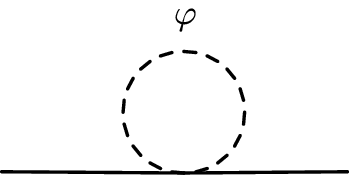}
\end{center}
contribute a divergent correction proportional to $\frac{m_{A}^{2}}{\epsilon}$. This extra divergence is canceled in the $A_{\mu}$ self-energy by diagrams of the form
\begin{center}
\includegraphics{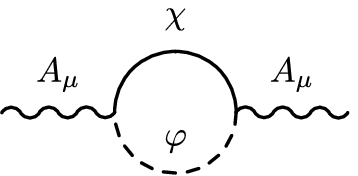}
\end{center}
However, since the $A_{5}$ is a scalar, unprotected by a gauge symmetry, this divergence is not eliminated in the final result. Excluding $A_{5}$, the low-energy theory was a 4-d Abelian Higgs model, and if not for $A_{5}$ the divergences in both phase would be the same. Therefore, the most direct route to ensuring that there are no problems subtracting infinities in the broken phase is to choose compactifications where the $A_{5}$ zero mode is absent. This is the case for $S^{1}/\mathbb{Z}_{2}$ orbifold compactifications. We found that when the theory is placed on $\mathbb{R}^{3,1} \times S^{1}/\mathbb{Z}_{2}$, the divergences in the symmetric and broken phases are the same. The decoupling of the heavy KK modes is then manifested.  

Without the $A_{5}$ zero mode the divergences in the LPEA can be consistently subtracted. We constructed RG-like equations for the scalar and gauge couplings with respect to the compactification scale $M$. The resulting solutions showed that the effective couplings decrease with $M$. Because the Higgs and gauge masses are proportional to the scalar and gauge couplings, respectively, these two also decrease with increasing scale $M$. The VEV also has has a scale dependence, but its effect on the scaling behavior of the masses is minimal. If $m_{\chi} > m_{A}$ at a low scale $M_{0}$, the values of the two masses will intersect at some higher scale. Were these the standard model Higgs and gauge bosons, the region with $m_{\chi} < m_{A}$ puts an upper limit on the compactification scale. The solutions for the scalar coupling also showed that even when $\lambda_{0} > 0$, $\lambda$ becomes negative at high scales. A negative $\lambda$ implies an unstable vacuum, which is physically unacceptable. Assuming that the effective field theory description is valid, this again places an upper limit on $M$. 

Constraining the masses and couplings to only physically acceptable regions could possibly be used to determine the compactification scale. In the future we will have to extend this analysis to a KK version of the electroweak model. Using the known physical constraints on Higgs and gauge masses, we might be able to learn something about the size of any extra dimensions that might exist. It may also be worthwhile to study heavy mode decoupling in the Higgs model with fermions. This would make our model more physically relevant, and the KK modes of the fermions may have a noticeable effect on the Peskin-Takeuchi parameters. The role of the chiral anomaly and it's interplay with KK mode decoupling is another outstanding problem that warrants further investigation. 

\renewcommand{\theequation}{A-\arabic{equation}}

 \setcounter{equation}{0}  

\appendix
\section{Light-Particle Effective Action in $\phi^{4}$ Theory}
\label{LPEAphi4Theory}
In this appendix we consider the light-particle effective action of the five dimensional action 
\begin{gather}
S
=
\int d^{5}\bar{x}
\left[
\bar{Z}_{\phi}
|\partial_{M} \bar{\phi}|^{2}
-
\bar{Z}_{\phi}
\bar{m}_{b}^{2}
|\bar{\phi}|^{2}
-
\frac{\bar{Z}_{\phi}^{2} \bar{\lambda}_{b}}{3!}
|\bar{\phi}|^{4}
\right].
\label{PhiFourAction}
\end{gather}
Here $\bar{\phi}$ denotes a complex scalar field in five dimensions, and has a field redefinition $\bar{Z}_{\phi}$. The coefficients $\bar{\lambda}_{b}$, and $\bar{m}_{b}^{2}$ are the bare couplings and we express them in terms of the physical couplings and their counter terms like so: 
\begin{gather}
\bar{\lambda}_{b} = \bar{\lambda} (1 + \bar{ \delta}_{\lambda}),
\qquad
\bar{m}_{b}^{2} = \bar{m}^{2} (1 + \bar{\delta}_{m^{2}}).
\end{gather}
Here, $\bar{\lambda}$ and $\bar{m}^{2}$ are the physical coupling and mass and $\bar{ \delta}_{\lambda}$ and $\bar{ \delta}_{m^{2}}$ are their corresponding counter terms. Since the $5$-th direction is compactified then $\bar{\phi}$ may be expanded in terms of a Fourier series: 
\begin{gather}
\bar{\phi}(\bar{x})
=
(2 \pi R)^{-1/2}
\sum_{n=-\infty}^{\infty}
\phi^{(n)}(x) e^{i n M \theta}
\label{FSeriesPhi}
\end{gather}
where $M = R^{-1}$. With the fields now represented as a Fourier series, the coordinate parametrizing the compact dimension $\theta$ can be integrated over, leaving us with a 4-d action for the KK mode fields $\phi^{(n)}(x)$.

We wish to find the LPEA of the 4-d theory using the ``$K$-Matrix'' method of Weisberger \cite{Weisberger:1981xe, Kazama:1981fx}. To explain this method, lets start by assuming a theory with a set of ``light'' fields $\{\phi_{i}\}$ and a set of ``heavy'' fields $\{\Phi_{\alpha}\}$. Here $\phi_{i}$ and $\Phi_{\alpha}$ denote light and heavy field types $i$ and $\alpha$, respectively. The dynamics of the fields $\{\phi_{i}\}$ and $\{\Phi_{\alpha}\}$ are determined by the action $S[\phi,\Phi]$. The partition function $Z[j,J]$ is defined as
\begin{gather}
Z[j,J]
=
- i \log \int [\D \phi_{i} \D \Phi_{\alpha}] e^{
i S[\phi,\Phi] +i \int ( j \cdot \phi + J \cdot \Phi)}.
\end{gather}
Here $j_{i}$ and $J_{\alpha}$ are classical sources for the fields $\phi_{i}$ and $\Phi_{\alpha}$, respectively. In order to get the proper low-energy effective field theory, define the {\it light-particle effective action} $\bar{\Gamma}$ as the Legendre transform of $Z$ with respect to only the light particle current:
\begin{gather}
\bar{\Gamma}[\phi_{c}]
=
Z[j,0]
-
\int j \cdot \phi_{c}.
\end{gather}
The functional $\bar{\Gamma}$ generates all diagrams that are 1PI with respect to the light fields $\phi_{i}$, but not 1PI with respect to the heavy fields $\Phi_{\alpha}$. The light particle effective action, therefore, includes all corrections to the couplings and masses from diagrams containing heavy internal loops. Before we can give the definition of the LPEA, we must first define the $K$-matrix:
\begin{gather}
K_{x,y;i,j}(\phi_{c})
= 
\left.
\frac{\delta^{2} S[\phi_{c},\Phi]}{\delta \Phi_{i}(x) \Phi_{j}(y)}
\right|_{\Phi = \Phi(\phi_{c})}.
\end{gather} 
Here $\phi_{c}$ is a set of classical values for the light particle fields and $\Phi(\phi_{c})$ denotes the classical solution to $\frac{\delta \Gamma}{\delta\Phi} = 0$ with $\phi = \phi_{c}$. Note that $\Gamma$ is the effective action that is 1PI in {\it both} the heavy and light fields. In practice we can approximate $\Phi(\phi_{c})$ by replacing $\Gamma$ with the classical action. With the $K$-matrix, the LPEA is defined as
\begin{gather}
\bar{\Gamma}[\phi_{c}]
=
S[\phi_{c},\Phi(\phi_{c})]
+
\frac{i}{2}
\Tr \log \left[
K(\phi_{c})
K^{-1}(0)
\right]
=
S[\phi_{c},\Phi(\phi_{c})]
+
\delta \bar{\Gamma}[\phi_{c}].
\end{gather}
We have suppressed indices for the sake of clarity. The trace in this context refers to the trace over everything: space-time position, particle type, group indices, KK modes, etc. Absent from this trace are the light particle types, since by definition the LPEA is obtained by integrating out only heavy particle species. For practical purposes the LPEA must be computed perturbatively. To obtain a definition of the LPEA that is more friendly to perturbative methods, we will split the $K$-matrix into two parts: a free field and an interaction piece. The free field part of the $K$-matrix is defined as 
\begin{gather}
K_{x,y;i,j}(0)
= 
\left.
\frac{\delta^{2} S_{0}[0,\Phi]}{\delta \Phi_{i}(x) \Phi_{j}(y)}
\right|_{\Phi = \Phi(0)}
\end{gather}
where $S_{0}$ is the free field part of the action. The interaction piece is defined as
\begin{gather}
\delta K_{x,y;i,j}(\phi_{c})
=
\left.
\frac{\delta^{2} S_{int}[\phi_{c},\Phi]}{\delta \Phi_{i}(x) \Phi_{j}(y)}
\right|_{\Phi = \Phi(\phi_{c})}
\end{gather}
where $S_{int}$ contains all the interaction terms of the action. Note that $K(\phi_{c}) = K(0) + \delta K(\phi_{c})$. If we assume that the couplings in the interaction piece of the action are small, then $\log K(\phi_{c}) K^{-1}(0)$ can be expanded in terms of the ``small'' interaction term $\delta K$. The definition for the perturbative LPEA correction is:  
\begin{gather}
\delta \bar{\Gamma}[\phi_{c}]
=
\frac{i}{2}\textrm{Tr}\log K(\phi_{c}) K^{-1}(0)
=
\frac{i}{2}\textrm{Tr}\log\left[1  + \delta K(\phi_{c}) K^{-1}(0)\right]
\nn \\
=
\frac{i}{2}\sum_{k=1}^{\infty}
\frac{(-1)^{k+1}}{k}\textrm{Tr}[(\delta K(\phi_{c}) K^{-1}(0))^{k}]
=
\frac{i}{2}\sum_{k=1}^{\infty}
\frac{(-1)^{k+1}}{k}\textrm{Tr}[(B(\phi_{c}))^{k}].
\end{gather}
We have defined a new matrix functional $B(\phi_{c})$ as $B(\phi_{c}) = \delta K(\phi_{c}) K^{-1}(0)$. In a Kaluza-Klein model, the dimensionally reduced action always involves an infinite number of KK modes, $\Phi_{i}^{(n)}$ where $n$ is the KK index. Assuming a high compactification scale compared to the zero mode masses, the light fields are the zero modes $\phi_{i} = \Phi^{(0)}_{i}$, and those fields with nonzero KK index are considered heavy. In this case it is a good idea to label KK indices of the $K$-Matrix explicitly:
\begin{gather}
K_{x,y;i,j}^{(n,m)}(\phi_{c})
=
\left.
\frac{\delta^{2} S[\phi_{c} ,\Phi]}{\delta \Phi_{i}^{(n)}(x) \Phi_{j}^{(-m)}(y)}
\right|_{\Phi = \Phi(\phi_{c})}.
\end{gather}
In each of the models we have considered $\Phi^{(n)}(\phi_{c})$ vanishes. This in turn leads to a vanishing of all off-diagonal $K$-matrix elements. This simplifies the perturbative expression for the LPEA so that the trace over KK modes is a single sum: 
\begin{gather}
\delta \bar{\Gamma}[\phi_{c}]
=
\frac{i}{2}
\sum_{\stackrel{n = -\infty}{n\neq 0}}^{\infty}
\sum_{k=1}^{\infty}
\frac{(-1)^{k+1}}{k}\textrm{Tr}[(B^{(n)}(\phi_{c}))^{k}]
=
i
\sum_{n = 0}^{\infty}
\sum_{k=1}^{\infty}
\frac{(-1)^{k+1}}{k}\textrm{Tr}[(B^{(n)}(\phi_{c}))^{k}].
\label{KKmodeLPEAExpansion}
\end{gather}
Here we have excluded the $n=0$ term in the sum over the KK tower states since we are only integrating out the heavy modes. The KK modes include negative indices because we assume an $S^{1}$ compactification. Had we chosen an orbifold compactification there would be no negative KK modes, and the final result in (\ref{KKmodeLPEAExpansion}) would still have the factor of $\frac{1}{2}$.

In what follows we will compute the effective action for two cases: the first case being when the vacuum respects the global $U(1)$ symmetry of the action (i.e. $\bar{m}^{2} >0$), and the second case when the vacuum does not respect $U(1)$ ($\bar{m}^{2}<0$).

\subsection{$\phi^{4}$ Without Spontaneous Symmetry Breaking}
Here we will consider the case when $\bar{m}^{2}>0$. With a positive mass squared the classical vacuum lies at $|\phi| = 0$. Using the Fourier series expansion for $\bar{\phi}$ (\ref{FSeriesPhi}) we can integrate over the 5-th direction. Doing so results in the tree-level action:
\begin{gather}
S=
\int d^{4}x
\left[
\sum_{n = -\infty}^{\infty}
\left(
\partial_{\mu} \phi^{(n)} \partial^{\mu}\phi^{\ast (-n)}
-
(m^{2} + n^{2} M^{2}) \phi^{(n)} \phi^{\ast (-n)} 
\right)
\right.
\nn \\
\left.
-
\frac{\lambda}{3!}
\sum_{n,m,k=-\infty}^{\infty}
\phi^{(n)}
\phi^{(m)}
\phi^{\ast (k)}
\phi^{\ast (-n-m-k)}
\right]
\end{gather}
where $m^{2} = \bar{m}^{2}$, $\lambda = \frac{\bar{\lambda}}{2 \pi R}$. Since the heavy mode sector will be integrated out in the low-energy theory, the only relevant counter terms will be those involving the light sector fields. The counter term action is therefore 
\begin{gather}
\delta S
=
\int d^{4}x
\Bigg[
\delta_{\phi} |\partial_{\mu} \phi|^{2}
-
m^{2} (\delta_{m^{2}} + \delta_{\phi}) |\phi|^{2} 
-
\frac{\lambda}{3!}
(\delta_{\lambda} + 2 \delta_{\phi})
|\phi|^{4}
\Bigg].
\end{gather}
In this appendix we will limit our investigation to the case when $m^{2} \ll M^{2}$ thereby making $\phi = \phi^{(0)}$ a low-energy degree of freedom. The classical equations of motion for the Fourier modes of $\bar{\phi}$ are:
\begin{gather}
\partial^{2} \phi^{(n)}
+
(m^{2} + n^{2} M^{2}) \phi^{(n)}
+
\frac{\lambda}{3}\sum_{m,k=-\infty}^{\infty}
\phi^{(m)} \phi^{(k)} \phi^{\ast (n-m -k)}
=
J \delta_{n,0}.
\end{gather}
With no external current for the heavy KK modes, the classical solution to the equations of motion are $\phi^{(n)} = 0$ for $n \neq 0$. The $B$-matrix is therefore
\begin{gather}
B^{(n)}(\phi)
=
\left(
\begin{array}{cc}
-  \frac{2 \lambda}{3}  \Delta_{\phi}^{(n)} |\phi|^{2} & - \frac{\lambda}{3}  \Delta_{\phi}^{(n)} \phi^{2}
\\
- \frac{\lambda}{3}  \Delta_{\phi}^{(n)} \phi^{\ast 2} & - \frac{2 \lambda}{3}  \Delta_{\phi}^{(n)} |\phi|^{2}
\end{array}
\right)
\end{gather}
where $\Delta_{\phi}^{(n)} = - (\partial^{2} + m^{2} + n^{2}M^{2})^{-1}$. Using the expansion (\ref{KKmodeLPEAExpansion}) for the LPEA, we find that
\begin{gather}
\delta \bar{\Gamma}[\phi]
=
-
\frac{4 i \lambda}{3} 
\sum_{n=1}^{\infty}
\textrm{Tr}\left[
\Delta_{\phi}^{(n)} |\phi|^{2}
\right]
-
\frac{5 i \lambda^{2}}{9}
\sum_{n=1}^{\infty}
\textrm{Tr}\left[
(\Delta_{\phi}^{(n)})^{2} |\phi|^{4}
\right]
+
\cdots
.
\label{LogExpansionSym}
\end{gather}
Here we have only gone to second order in the $B$-matrix since higher orders only lead to convergent corrections to irrelevant operators. Below is a list of the divergent corrections to the $\phi$ self-energy and quartic coupling.

\begin{itemize}
\item {\bf $\phi$ Self-Energy Operator:}
\begin{gather}
-
\int \Sigma_{\phi}(p^{2}) |\phi|^{2}
=
-
\frac{4 i \lambda}{3}
\sum_{n=1}^{\infty}
\Tr\left[\Delta_{\phi}^{(n)} |\phi|^{2} \right]
\end{gather}
The divergent and $M$ dependent corrections to the $\phi$ self-energy correction are
\begin{gather}
\Sigma_{\varphi}(p^{2})
=
\frac{4 i \lambda}{3}
\sum_{n=1}^{\infty}
A_{0}^{(n)}(m^{2})
=
\frac{\lambda m^{2}}{12 \pi^{2} \epsilon}
\left(
\frac{\mu}{M}
\right)^{\epsilon}
+
\frac{\lambda M^{2} \zeta(3)}{24 \pi^{4}}
\end{gather}
$A_{0}^{(n)}$ is a modified version of the first Passarino-Veltman function \cite{Passarino:1978jh} defined in appendix \ref{KKModeSumsAppendix}.

\item {\bf $|\phi|^{4}$ Vertex Operator:}
\begin{gather}
-
\frac{\Gamma_{|\phi|^{4}}}{3!} \int |\phi|^{4}
=
- \frac{5 i \lambda^{2}}{9}
\sum_{n=1}^{\infty}
\Tr\left[
(\Delta_{\phi}^{(n)})^{2}
|\phi|^{4}
\right]
\end{gather}
The divergent part is
\begin{gather}
\Gamma_{|\phi|^{4}}
=
\frac{10 i \lambda^{2}}{3}
\sum_{n=1}^{\infty}
B_{0}^{(n)}(p^{2};m^{2},m^{2})
=
\frac{5 \lambda^{2}}{24 \pi^{2} \epsilon}
\left(
\frac{\mu}{M}
\right)^{\epsilon}
\end{gather}
$B_{0}^{(n)}$ is a modified version of the second Passarino-Veltman function defined in appendix \ref{KKModeSumsAppendix}. 
\end{itemize}

\subsubsection{Subtraction of the Mass and Coupling Divergences}
The finiteness of the $\phi$ self-energy requires that the mass counter term and field redefinition satisfy:
\begin{gather}
\Sigma(p^{2})
-
\delta_{\phi}
p^{2}
+
(\delta_{m^{2}} + \delta_{\phi})
m^{2}
=0.
\end{gather}
This implies that
\begin{gather}
\delta_{\phi}
=0
,
\quad
\delta_{m^{2}}
=
-
\frac{\lambda}{12 \pi^{2} \epsilon}
\left(
\frac{\mu}{M}
\right)^{\epsilon}
-
\frac{\lambda M^{2} \zeta(3)}{24 \pi^{4} m^{2}}.
\end{gather}
Finiteness of the $\phi$ 4-point vertex correction requires
\begin{gather}
\Gamma_{|\phi|^{4}}
+
(\delta_{\lambda} + 2 \delta_{\phi})
\lambda
=0,
\end{gather}
which implies that
\begin{gather}
\delta_{\lambda}
=
-
\frac{5 \lambda}{24 \pi^{2} \epsilon}
\left(
\frac{\mu}{M}
\right)^{\epsilon}.
\end{gather}
This result should be compared to result for $\delta_{\lambda}$ in the Abelian Higgs model (\ref{CT2}). Note that if one sets $e = 0$ in (\ref{CT2}) the results for $\delta_{\lambda}$ are the same.

In our subtraction scheme we subtract the divergent pole $\frac{1}{\epsilon}$ and the finite $\log \mu/M$ part from the loop corrections. This is to ensure that the final result for the one-loop corrected coupling (including the counter term) is not dependent on $M$. The bare coupling is therefore
\begin{gather}
\lambda_{b} 
=
\lambda (1 + \delta_{\lambda}) 
=
\lambda
\left(
1
-
\frac{5 \lambda}{24 \pi^{2} \epsilon}
\left(
\frac{\mu}{M}
\right)^{\epsilon}
\right)
\rightarrow
\lambda\left( 1 -  \frac{5 \lambda}{24 \pi^{2} \epsilon} + \frac{5 \lambda}{24 \pi^{2} } \log \left[\frac{M}{\mu}\right] \right).
\label{LambdaCorrected}
\end{gather}
We can define a RG-like equation for the coupling with respect to the compactification scale $M$. Keep in mind that the bare coupling is an intrinsic parameter of the theory, and so it should remain fixed with respect to the scale $M$.  Differentiating (\ref{LambdaCorrected}) with respect to $\log M$, and keeping terms to leading order we find that
\begin{gather}
\frac{d \lambda }{d \log M}
= 
-
\frac{5 \lambda^{2}}{24 \pi^{2}}.
\end{gather}
The sign on the right hand side indicates that the solution for $\lambda$ decreases as the scale $M$ increases. Unlike the Abelian Higgs model, the anomalous dimension of the scalar field vanishes at the one-loop order, so the VEV does not change with $M$. The leading order scaling behavior of the Higgs mass will be same as $\lambda$.

\subsection{$\phi^{4}$ with Spontaneous Symmetry Breaking}
Here we will outline the derivation of the LPEA in the case of spontaneous symmetry breaking. Taking $\bar{m}^{2}<0$ the global $U(1)$ symmetry is broken in the vacuum state of the theory. Classically, the vacuum is now located at $|\bar{\phi}| = \frac{\bar{v}}{\sqrt{2}}$ where $\bar{v}^{2} = - \frac{6 \bar{m}^{2}}{\bar{\lambda}}$. Expanding $\bar{\phi}$ around the classical vacuum:
\begin{gather}
\bar{\phi} = \frac{\bar{\varphi} + i \bar{\chi} + i \bar{v}}{\sqrt{2}},
\end{gather}
and substituting this for the field $\bar{\phi}$ into the action (\ref{PhiFourAction}) we obtain
\begin{gather}
S
=
\int d^{5}\bar{x}
\left[
\frac{\bar{Z}_{\phi}}{2}
\partial_{M} \bar{\varphi}
\partial^{M} \bar{\varphi}
+
\frac{\bar{Z}_{\phi}}{2}
\partial_{M} \bar{\chi}
\partial^{M} \bar{\chi}
-
\frac{\bar{Z}_{\phi}}{12} (\bar{Z}_{\phi} \bar{\lambda}_{b} \bar{v}^{2} + 6 \bar{m}_{b}^{2}) \bar{\varphi}^{2}
-
\frac{\bar{Z}_{\phi}}{4}
(\bar{Z}_{\phi} \bar{\lambda}_{b} \bar{v}^{2} + 2 \bar{m}_{b}^{2}) \bar{\chi}^{2}
\right.
\nn \\
\left.
-
\frac{\bar{Z}_{\phi}}{6}
(\bar{Z}_{\phi} \bar{\lambda}_{b} \bar{v}^{2} + 6 \bar{m}_{b}^{2})
\bar{v}
\bar{\chi}
-
\frac{\bar{Z}_{\phi}^{2} \bar{\lambda}_{b} \bar{v}}{3!}
\bar{\chi}
(\bar{\varphi}^{2} + \bar{\chi}^{2})
-
\frac{\bar{Z}_{\phi}^{2} \bar{\lambda}_{b}}{4!}
(\bar{\varphi}^{2} + \bar{\chi}^{2})^{2}
\right].
\end{gather}
Once we integrate out the 5-th dimension, the tree-level action becomes
\begin{gather}
S
=
\int d^{4}x
\left[
\sum_{n = -\infty}^{\infty}
\left(
\frac{1}{2} \partial_{\mu} \varphi^{(n)} \partial^{\mu}\varphi^{(-n)}
+
\frac{1}{2} \partial_{\mu} \chi^{(n)} \partial^{\mu}\chi^{(-n)}
-
\frac{1}{2} n^{2} M^{2} \varphi^{(n)} \varphi^{(-n)} 
\right.
\right.
\nn \\
\left.
-
\frac{1}{2} \left[m_{\chi}^{2} + n^{2} M^{2}\right] \chi^{(n)} \chi^{(-n)} 
\right)
-
\frac{\eta}{3!}
\sum_{k,l=-\infty}^{\infty}
(\varphi^{(k)} \varphi^{(l)} + \chi^{(k)} \chi^{(l)})\chi^{(-k-l)}
\nn \\
-
\frac{\lambda}{4!}
\sum_{k,l,n=-\infty}^{\infty}
\left(
\varphi^{(k)} \varphi^{(l)} \varphi^{(n)} \varphi^{(-k-l-n)} 
+ 2 \chi^{(k)} \chi^{(l)} \varphi^{(n)} \varphi^{(-k-l-n)} + \chi^{(k)} \chi^{(l)} \chi^{(n)} \chi^{(-k-l-n)} 
\right)
\Bigg]
\end{gather}
where
\begin{gather}
m_{\chi}^{2} = \frac{\bar{\lambda} \bar{v}^{2}}{3}
,
\qquad
v = \sqrt{2 \pi R} \, \bar{v}
,
\qquad
\eta = \frac{\bar{\lambda} \bar{v}}{\sqrt{2 \pi R}}
,
\qquad
\lambda = \frac{\bar{\lambda}}{2 \pi R}.
\end{gather}
Since the zero mode of the Goldstone boson $\varphi = \varphi^{(0)}$ is massless then it will always be present in the low-energy effective action. Since the zero mode of the Higgs field $\chi = \chi^{(0)}$ in general has a mass, there are two scenarios to consider: $1 \textrm{ TeV} \ll m \sim M$ and $1 \textrm{ TeV} \sim m \ll M$. The first scenario has a trivial low-energy action, consisting of only the zero mode $\varphi$. It should be expected in light of \cite{Weisberger:1981xe} that the LPEA of the Goldstone field $\varphi$ alone should be that of a free massless scalar. As a consistency check it can be easily shown that this is the case to one-loop order.

Since the heavy mode sector will be integrated out in the low-energy limit, the only relevant counter terms will be those involving the light sector fields. The counter term action is therefore 
\begin{gather}
\delta S
=
\int d^{4}x
\Bigg[
\frac{\delta_{\phi}}{2} \partial_{\mu} \varphi \partial^{\mu}\varphi
+
\frac{\delta_{\phi}}{2} \partial_{\mu} \chi \partial^{\mu}\chi
-
\frac{\delta \sigma}{2 v}  \varphi^{2} 
-
\frac{1}{2} \left(m_{\chi}^{2} [\delta_{\lambda} + 2 \delta_{\phi}]+ \frac{\delta \sigma}{v} \right) \chi^{2} 
\nn \\
-
\delta \sigma \chi
-
\frac{\eta}{3!}
(\delta_{\lambda} + 2 \delta_{\phi})
\chi
(\varphi^{2} + \chi^{2})
-
\frac{\lambda}{4!}
(\delta_{\lambda} + 2 \delta_{\phi})
(\varphi^{2} + \chi^{2})^{2}
\Bigg]
\end{gather}
where 
\begin{gather}
\delta_{m^{2}}
=
\bar{\delta}_{m^{2}}
,
\qquad
\delta_{\lambda} = \bar{\delta}_{\lambda}
,
\qquad
\frac{\delta \sigma}{v} = \frac{\lambda v^{2}}{6} (\delta_{\lambda} + \delta_{\phi} - \delta_{m^{2}}).
\end{gather}
By definition, the $B$-matrix in this case is
\begin{gather}
B^{(n)}(\varphi,\chi)
=
\left(
\begin{array}{cc}
-\Delta_{\varphi}^{(n)} (\frac{\eta}{3}\chi + \frac{\lambda}{2}\varphi^{2}+\frac{\lambda}{6} \chi^{2}) & -\Delta_{\chi}^{(n)}  (\frac{\eta}{3}\varphi + \frac{\lambda}{3}\varphi \chi) 
\\
-\Delta_{\varphi}^{(n)}  (\frac{\eta}{3}\varphi + \frac{\lambda}{3}\varphi \chi) & -\Delta_{\chi}^{(n)}  (\eta \chi + \frac{\lambda}{2}\chi^{2}+\frac{\lambda}{6} \varphi^{2})
\end{array}
\right)
\end{gather}
where $\Delta_{\varphi}^{(n)} = - (\partial^{2} + n^{2} M^{2})^{-1}$ and $\Delta_{\chi}^{(n)} = - (\partial^{2} + m_{\chi}^{2} + n^{2} M^{2})^{-1}$. Using the formula given in (\ref{KKmodeLPEAExpansion}), the correction to the LPEA is
\begin{gather}
\delta \bar{\Gamma}[\varphi,\chi]
=
i
\sum_{n=1}^{\infty}
\textrm{Tr}\left[
-\Delta_{\varphi}^{(n)} (\frac{\eta}{3}\chi + \frac{\lambda}{2} \varphi^{2} +\frac{\lambda}{6} \chi^{2})
-
\Delta_{\chi}^{(n)} (\eta \chi + \frac{\lambda}{2}\chi^{2} + \frac{\lambda}{6} \varphi^{2})
\right]
\nn \\
-
\frac{i}{2}
\sum_{n=1}^{\infty}
\textrm{Tr}\Bigg[\left[\Delta_{\varphi}^{(n)} (\frac{\eta}{3}\chi + \frac{\lambda}{2} \varphi^{2} +\frac{\lambda}{6} \chi^{2})\right]^{2}
+
\left[ \Delta_{\chi}^{(n)}  (\eta \chi + \frac{\lambda}{2}\chi^{2} + \frac{\lambda}{6} \varphi^{2}) \right]^{2}
\nn \\
+
2
\Delta_{\varphi}^{(n)} \left(
\frac{\eta}{3} \varphi + \frac{\lambda}{3} \chi \varphi
\right)
\Delta_{\chi}^{(n)} 
\left(
\frac{\eta}{3} \varphi + \frac{\lambda}{3} \chi \varphi
\right)
\Bigg]
+\cdots.
\label{LogExpansionBroken}
\end{gather}
In the following, all of the divergent loop corrections are calculated. Please note that we have only included those Passarino-Veltman functions which contain divergences.

\begin{itemize}
\item {\bf $\chi$ One-Point Vertex Operator:}
\begin{gather}
\sigma
=
\frac{i\eta}{3}
\sum_{n=1}^{\infty}
\left[
A_{0}^{(n)}(0)
+
3 A_{0}^{(n)}(m_{\chi}^{2})
\right]
=
\frac{\lambda^{2} v^{3}}{48 \pi^{2} \epsilon}
\left(
\frac{\mu}{M}
\right)^{\epsilon}
+
\frac{\lambda  v M^{2} \zeta(3)}{24 \pi^{4}}
\end{gather}

\item {\bf $\varphi$ Self-Energy Operator:}

\begin{gather}
\Pi_{\varphi}
=
\frac{i \lambda}{3}
\sum_{n=1}^{\infty}
\left[
3 A_{0}^{(n)}(0)
+
A_{0}^{(n)}(m_{\chi}^{2})
+
2 m_{\chi}^{2}
B_{0}^{(n)}(p^{2} ; 0 , m_{\chi}^{2})
\right]
\nn \\
=
\frac{\lambda m_{\chi}^{2}}{16 \pi^{2} \epsilon}
\left(
\frac{\mu}{M}
\right)^{\epsilon}
+
\frac{\lambda M^{2} \zeta(3)}{24 \pi^{4}}
\end{gather}

\item {\bf $\chi^{2}$ Self-Energy Operator:}

\begin{gather}
\Sigma(p^{2})
=
\frac{i \lambda}{3} 
\sum_{n=1}^{\infty}
\bigg[
A_{0}^{(n)}(0)
+
3 A_{0}^{(n)}(m_{\chi}^{2})
+
m_{\chi}^{2} B_{0}^{(n)}(p^{2};0,0)
+
9 m_{\chi}^{2} B_{0}^{(n)}(p^{2};m_{\chi}^{2},m_{\chi}^{2})
\bigg]
\nn \\
=
\frac{13 \lambda m_{\chi}^{2}}{48 \pi^{2} \epsilon}
\left(
\frac{\mu}{M}
\right)^{\epsilon}
+
\frac{\lambda M^{2} \zeta(3)}{24 \pi^{4}}
\end{gather}


\item {\bf The $\chi \varphi^{2}$ Operator:}

\begin{gather}
\Gamma_{\chi \varphi^{2}}
=
\frac{i \eta \lambda}{3}
\sum_{n=1}^{\infty}
\Bigg[
3 B_{0}^{(n)}(p^{2};0,0)
+
3
B_{0}^{(n)}(p^{2};m_{\chi}^{2},m_{\chi}^{2})
+
4
B_{0}^{(n)}(p^{2};0,m_{\chi}^{2})
\Bigg]
\nn \\
=
\frac{5 \lambda \eta}{24 \pi^{2}\epsilon}
\left(
\frac{\mu}{M}
\right)^{\epsilon}
\end{gather}

\item {\bf $\chi^{3}$ Vertex Operator:}

\begin{gather}
\Gamma_{\chi^{3}}
=
\frac{i \eta \lambda}{3}
\sum_{n=1}^{\infty}
\Bigg[
B_{0}^{(n)}(p^{2};0,0)
+
9
B_{0}^{(n)}(p^{2};m_{\chi}^{2},m_{\chi}^{2})
\Bigg]
=
\frac{5 \lambda \eta}{24 \pi^{2} \epsilon}
\left(
\frac{\mu}{M}
\right)^{\epsilon}
\end{gather}

\item {\bf $\varphi^{2} \chi^{2}$ Vertex Operator:}

\begin{gather}
\Gamma_{\varphi^{2} \chi^{2}}
=
\frac{i \lambda^{2}}{3}
\sum_{n=1}^{\infty}
\Bigg[
3 B_{0}^{(n)}(p^{2};0,0)
+
3 B_{0}^{(n)}(p^{2};m_{\chi}^{2},m_{\chi}^{2})
+
4 B_{0}^{(n)}(p^{2};0,m_{\chi}^{2})
\Bigg]
\nn \\
=
\frac{5 \lambda^{2}}{24 \pi^{2} \epsilon}
\left(
\frac{\mu}{M}
\right)^{\epsilon}
\end{gather}

\item {\bf $\varphi^{4}$ Vertex Operator:}

\begin{gather}
\Gamma_{\varphi^{4}}
=
\frac{i \lambda^{2}}{3}
\sum_{n=1}^{\infty}
\Bigg[
9
B_{0}^{(n)}(p^{2};0,0)
+
B_{0}^{(n)}(p^{2};m_{\chi}^{2},m_{\chi}^{2})
\Bigg]
=
\frac{5 \lambda^{2}}{24 \pi^{2}\epsilon}
\left(
\frac{\mu}{M}
\right)^{\epsilon}
\end{gather}

\item {\bf $\chi^{4}$ Vertex Operator:}

\begin{gather}
\Gamma_{\chi^{4}}
=
\frac{ i \lambda^{2}}{3}
\sum_{n=1}^{\infty}
\Bigg[
B_{0}^{(n)}(p^{2};0,0)
+
9
B_{0}^{(n)}(p^{2};m_{\chi}^{2},m_{\chi}^{2})
\Bigg]
=
\frac{5 \lambda^{2}}{24 \pi^{2} \epsilon}
\left(
\frac{\mu}{M}
\right)^{\epsilon}
\end{gather}
\end{itemize}

\subsubsection{Subtraction of the Mass and Coupling Divergences}
The conditions for finiteness of the LPEA are:
\begin{gather}
\sigma + \delta \sigma
=
0,
\label{vevCancellation}
\\
\Pi_{\varphi}(p^{2})
-
\delta_{\phi}
p^{2}
+
\frac{\delta \sigma}{v}
=0,
\\
\Sigma(p^{2})
-
\delta_{\phi}
p^{2}
+
(\delta_{\lambda} + 2 \delta_{\phi})
m_{\chi}^{2}
+
\frac{\delta \sigma}{v}
=0,
\\
\Gamma_{\chi \varphi^{2}}
+
(\delta_{\lambda} + 2 \delta_{\phi}) \eta
=0
,
\quad
\Gamma_{\chi^{3}}
+
(\delta_{\lambda} + 2 \delta_{\phi}) \eta
=0,
\\
\Gamma_{\varphi^{2} \chi^{2}}
+
(\delta_{\lambda} + 2 \delta_{\phi}) \lambda
=0
,
\quad
\Gamma_{\varphi^{4}}
+
(\delta_{\lambda} + 2 \delta_{\phi}) \lambda
=0
,
\quad
\Gamma_{\chi^{4}}
+
(\delta_{\lambda} + 2 \delta_{\phi}) \lambda
=0.
\end{gather}
All of these equations are satisfied if
\begin{gather}
\frac{\delta \sigma}{v} 
=
-
\frac{\lambda^{2} v^{2}}{48 \pi^{2} \epsilon}
\left(
\frac{\mu}{M}
\right)^{\epsilon}
-
\frac{\lambda  M^{2} \zeta(3)}{24 \pi^{4}}
,
\qquad
\delta_{\phi}
=0
,
\qquad
\delta_{\lambda}
=
-
\frac{5 \lambda}{24 \pi^{2} \epsilon}
\left(
\frac{\mu}{M}
\right)^{\epsilon}.
\end{gather}
Note that there is no mass shift for the Goldstone at $p^{2} = 0$, which implies that the Goldstone remains massless to one-loop order. The $m^{2}$ counter term can be found by the relation $\frac{\delta \sigma}{v} = m^{2} (\delta_{m^{2}} -  \delta_{\lambda} - \delta_{\phi})$, which when inverted to find $\delta_{m^{2}}$ becomes:
\begin{gather}
\delta_{m^{2}}
=
\frac{\delta \sigma}{v m^{2}}
+
\delta_{\lambda} + \delta_{\phi}.
\end{gather}
Thus we find that
\begin{gather}
\delta_{m^{2}} m^{2}
=
-\frac{\lambda}{12 \pi^{2} \epsilon}
\left(
\frac{\mu}{M}
\right)^{\epsilon}
-
\frac{\lambda  M^{2} \zeta(3)}{24 \pi^{4}}.
\end{gather}
This result should be compared to the symmetric phase ($\bar{m}^{2}>0$). The corrections to the mass and coupling parameter are the same in both phases, and therefore the same divergences are shared between the two phases.

\renewcommand{\theequation}{B-\arabic{equation}}

 \setcounter{equation}{0}  

\section{Zeta Function Regularization}
Thoughout this paper we have had to deal with divergent sums of the form
\begin{gather}
\sum_{n=1}^{\infty} \log\left[
\frac{m^{2} + n^{2} M^{2}}{\mu^{2}}
\right]
,
\qquad
\sum_{n=1}^{\infty} (m^{2} + n^{2} M^{2}) \log\left[
\frac{m^{2} + n^{2} M^{2}}{\mu^{2}}
\right].
\label{Sums}
\end{gather}
Although these series are divergent, some sense can still be made of them. Consider the Riemann zeta function $\zeta(s)$. For the domain $s > 1$ the zeta function can be written as an infinite series:
 \begin{gather}
 \zeta(s)
 =
 \sum_{n=1}^{\infty}
 n^{-s}.
 \end{gather}
Although this series is divergent when $s < 1$, the function $\zeta(s)$ nevertheless has a unique analytic continuation onto the entire complex plane. For our proposes let us define a generalized zeta function $\zeta_{L}(s)$ as
\begin{gather}
\zeta_{L}(s)
=
\sum_{n=1}^{\infty}
[ m^{2} + n^{2} M^{2} ]^{-s}
\end{gather}
for $s >1/2$. For $s < 1/2$ the series is divergent. However, like the Riemann zeta function, we can show that $\zeta_{L}(s)$ also has a unique analytic continuation. Note that the derivative of this function is
\begin{gather}
\zeta^{\prime}(s)
=
-
\sum_{n=1}^{\infty}
(m^{2} + n^{2} M^{2} )^{-s}
\log\left[
m^{2} + n^{2} M^{2}
\right].
\end{gather}
Therefore the divergent sums in (\ref{Sums}) can be written as
\begin{gather}
-\zeta^{\prime}_{L}(-k)
-
\zeta_{L}(-k)
\log \mu^{2}
=
\sum_{n=1}^{\infty} 
(m^{2} + n^{2} M^{2})^{k}
\log\left[
\frac{m^{2} + n^{2} M^{2}}{\mu^{2}}
\right].
\end{gather}
Once we have obtained an expression for the analytic continuation of $\zeta_{L}$ we can use the expression above to define the divergent series. In order to accomplish this, we expand $(m^{2} +n^{2} M^{2})^{-s}$ using the binomial theorem. The generalized zeta function can now be written as
\begin{gather}
\zeta_{L}(s)
=
\sum_{n=1}^{\infty}
\sum_{k=0}^{\infty}
(-1)^{k}
\frac{\Gamma(s+k)}{\Gamma(k+1) \Gamma(s)}
m^{2 k}
(n M)^{-2 s -2 k}
\nn \\
=
M^{-2 s}
\sum_{k=0}^{\infty}
\alpha(s,k)
\zeta(2s+ 2 k)
\left(-\frac{m^{2}}{M^{2}}\right)^{k}.
\end{gather}
It is implied that if $\alpha$ is undefined at some value of $s$, the limit is taken if it exists. Note that we have used the definition of the Riemann zeta function to give the sum over $n$ a well defined result. The derivative of $\zeta_{L}$ is also important and it is given by:
\begin{gather}
\zeta_{L}^{\prime}(s)
=
M^{-2 s}
\sum_{k=0}^{\infty}
\left(-\frac{m^{2}}{M^{2}}\right)^{k}
\left[
\beta(s,k)
\zeta(2s+ 2 k)
+
\alpha(s,k)
\left(
2 
\zeta^{\prime}(2 s + 2 k)
-
\log M^{2} \zeta(2s+2k)
\right)
\right].
\end{gather}
Here $\beta$ is defined as
\begin{gather}
\beta(s,k)
=
\frac{\Gamma^{\prime}(s+k)\Gamma(s) - \Gamma(s+k) \Gamma^{\prime}(s)}{\Gamma(k+1)\Gamma^{2}(s)}.
\end{gather}
Note that if $s = 0,-1,-2,\dots$
\begin{align}
\alpha(s,k)
=
(-1)^{k}
\left(
\begin{array}{c}
-s
\\
k
\end{array}
\right)
\quad
&\textrm{for } \, k \leq -s
\\
\alpha(s,k)=0
\quad
&\textrm{for } \, k > -s
\\
\beta(0,0)
=
0
,
\quad
\beta(0,k) = \frac{1}{k}
\quad
&\textrm{for } \, k \geq 1
\\
\beta(-1,0)
=
0
,
\quad
\beta(-1,1)
=
1
,
\quad
\beta(-1,k) = -\frac{1}{k (k - 1)}
\quad
&\textrm{for } \, k \geq 2.
\end{align}
Since $\alpha$ vanishes when $k >-s$ the sum over $k$ in $\zeta_{L}$ truncates, and is thus trivially convergent. The derivative $\zeta^{\prime}_{L}$ has a finite radius of convergence with respect to the ratio $m^{2}/M^{2}$. The $\beta$ terms in the series comprise an alternating series which is convergent so long as the terms satisfy
\begin{gather}
\beta(s,k+1)
\zeta(2s+2k +2)
\left(\frac{m^{2}}{M^{2}}\right)^{k+1} 
\leq 
\beta(s,k)
\zeta(2s+2k)
\left(\frac{m^{2}}{M^{2}}\right)^{k} 
\quad
\textrm{for all } k
\\
\lim_{k\rightarrow \infty}
\beta(s,k)
\zeta(2s+2k)
\left(\frac{m^{2}}{M^{2}}\right)^{k} 
=0.
\end{gather}
Note that these are satisfied only if $m \leq M$. Fortunately, in this paper we are assuming that all zero mode masses are much smaller then the compactification mass, so we can rest assured that the sum over $k$ is convergent. If it is the case that $m > M$, the sum can be analytically continued to an entire function on the complex plane using the identity
\begin{gather}
\sum_{k=1}^{\infty}
\zeta(2 k) (-x)^{k-1}
=
\frac{\pi \coth(\pi \sqrt{x})}{2 \sqrt{x}}
-
\frac{1}{2 x}.
\end{gather}
We now have all the tools need to evaluate the divergent sums (\ref{Sums}):
\begin{gather}
\sum_{n=1}^{\infty}
\log\left[
\frac{m^{2} + n^{2} M^{2}}{\mu^{2}}
\right]
=
-
\frac{1}{2}
\log\left[
\frac{M^{2}}{4 \pi^{2} \mu^{2}}
\right]
+
\log\left[
\frac{\sinh(\pi \rho)}{\pi \rho}
\right]
\label{FirstSum}
\end{gather}
\begin{gather}
\sum_{n=1}^{\infty}
(m^{2} + n^{2} M^{2}) \log\left[
\frac{m^{2} + n^{2} M^{2}}{\mu^{2}}
\right]
=
-
\frac{m^{2}}{2}
\log\left[
\frac{M^{2}}{4 \pi^{2} \mu^{2}}
\right]
-
\frac{m^{2}}{2}
+
\frac{M^{2} \zeta(3)}{2 \pi^{2}}
\nn \\
-
m^{2}
\left(
\log\left[
-
2 \pi \rho e^{\frac{2 \pi \rho}{3} - \frac{1}{2}}
\right]
+
\frac{\zeta(3)
-
\textrm{Li}_{3}(e^{2 \pi \rho})
+
2 \pi \rho \textrm{Li}_{2}(e^{2 \pi \rho})
}{2 \pi^{2} \rho^{2}}
\right)
\label{SecondSum}
\end{gather}
where $\rho = \frac{m}{M}$. In the limit that we are considering, $m \ll M$. Therefore, the last term in (\ref{FirstSum}) and the term in parentheses in (\ref{SecondSum}) are both subleading compared to the other terms. For this reason these terms are ignored in our analysis.

\renewcommand{\theequation}{C-\arabic{equation}}

 \setcounter{equation}{0}  

\section{Common Kaluza-Klein Mode Sums}
\label{KKModeSumsAppendix}
In this paper we define two slightly modified versions of the PV functions:
\begin{gather}
A_{0}^{(n)}(m^{2})
=
\int \frac{d^{d}k}{(2 \pi)^{d}}
\frac{1}{k^{2} - m^{2} - n^{2} M^{2}},
\\
B_{0}^{(n)}(p^{2};m^{2},m^{\prime 2})
=
\int \frac{d^{d} k}{(2 \pi)^{d}}
\frac{1}{[k^{2} - m^{2} - n^{2} M^{2}] [(k+p)^{2} - m^{\prime 2} - n^{2} M^{2}]}.
\end{gather}
These two PV functions are all that are needed to evaluate the loop corrections in the models discussed in this paper. Summing over the KK modes using zeta function regularization we arrive at
\begin{gather}
\sum_{n=1}^{\infty}
A_{0}^{(n)}(m^{2})
\sim
-
\frac{i m^{2}}{16 \pi^{2} \epsilon}
\left(
\frac{\mu}{M}
\right)^{\epsilon}
-
\frac{i M^{2} \zeta(3)}{32 \pi^{4}},
\\
\sum_{n=1}^{\infty}
B_{0}^{(n)}(p^{2};m^{2},m^{\prime 2})
\sim
-
\frac{i}{16 \pi^{2} \epsilon}
\left(
\frac{\mu}{M}
\right)^{\epsilon},
\\
\sum_{n=1}^{\infty}
n^{2} M^{2} B_{0}^{(n)}(p^{2};m^{2},m^{\prime 2})
\sim
-
\frac{i M^{2} \zeta(3)}{32 \pi^{4}}.
\end{gather}

\bibliographystyle{h-elsevier}
\bibliography{KK-Decoupling-Article-Bib}

\end{document}